\documentclass[journal]{IEEEtran}

\usepackage{amssymb}
\usepackage{amsmath}
\usepackage{amsfonts}
\usepackage{multirow}
\usepackage{graphicx}
\usepackage{textcomp}
\usepackage{amsthm}
\renewcommand{\vec}[1]{\boldsymbol{#1}}
\usepackage{setspace}
\usepackage{subfigure}
\usepackage{algorithmic}
\usepackage[ruled,vlined,boxed,linesnumbered]{algorithm2e}
\usepackage[dvipsnames]{xcolor}
\usepackage{bm}
\usepackage{hyperref}

\allowdisplaybreaks[4]

\ifCLASSINFOpdf
\else
\fi

\begin{document}

\title{MTMD: Multi-Scale Temporal Memory Learning and Efficient Debiasing Framework for Stock Trend Forecasting}

\author{Mingjie Wang,
    Juanxi Tian,
    Mingze Zhang,
    Jianxiong Guo,~\IEEEmembership{Member,~IEEE},
    and Weijia Jia,~\IEEEmembership{Fellow,~IEEE}
	\thanks{Mingjie Wang and Juanxi Tian are with the Guangdong Key Lab of AI and Multi-Modal Data Processing, Department of Computer Science, BNU-HKBU United International College, Zhuhai 519087, China. (E-mail: mjwang0606@gmail.com; r130034034@mail.uic.edu.cn)

    Mingze Zhang is with the School of Engineering and Applied Sciences, Gonzaga University, WA 99258, USA. (E-mail: E15503075646@hotmail.com)
 
	Jianxiong Guo and Weijia Jia are with the Advanced Institute of Natural Sciences, Beijing Normal University, Zhuhai 519087, China, and also with the Guangdong Key Lab of AI and Multi-Modal Data Processing, BNU-HKBU United International College, Zhuhai 519087, China. (E-mail: jianxiongguo@bnu.edu.cn; jiawj@bnu.edu.cn)
	
	\textit{(Corresponding author: Jianxiong Guo; Weijia Jia.)}
	}
	\thanks{2471-285X \copyright\, 2025 IEEE. Personal use is permitted, but republication/redistribution requires IEEE permission.}}

\markboth{IEEE Transactions on Emerging Topics in Computational Intelligence}%
{Shell \MakeLowercase{\textit{et al.}}: A Sample Article Using IEEEtran.cls for IEEE Journals}


\maketitle

\begin{abstract}
The endeavor of stock trend forecasting is principally focused on predicting the future trajectory of the stock market, utilizing either manual or technical methodologies to optimize profitability. Recent advancements in machine learning technologies have showcased their efficacy in discerning authentic profit signals within the realm of stock trend forecasting, predominantly employing temporal data derived from historical stock price patterns. Nevertheless, the inherently volatile and dynamic characteristics of the stock market render the learning and capture of multi-scale temporal dependencies and stable trading opportunities a formidable challenge. This predicament is primarily attributed to the difficulty in distinguishing real profit signal patterns amidst a plethora of mixed, noisy data. In response to these complexities, we propose a Multi-Scale Temporal Memory Learning and Efficient Debiasing (MTMD) model. This innovative approach encompasses the creation of a learnable embedding coupled with external attention, serving as a memory module through self-similarity. It aims to mitigate noise interference and bolster temporal consistency within the model. The MTMD model adeptly amalgamates comprehensive local data at each timestamp while concurrently focusing on salient historical patterns on a global scale. Furthermore, the incorporation of a graph network, tailored to assimilate global and local information, facilitates the adaptive fusion of heterogeneous multi-scale data. Rigorous ablation studies and experimental evaluations affirm that the MTMD model surpasses contemporary state-of-the-art methodologies by a substantial margin in benchmark datasets. The source code can be found at \url{https://github.com/MingjieWang0606/MDMT-Public}.
\end{abstract}


\IEEEpeerreviewmaketitle

\section{Introduction}

\IEEEPARstart{F}{inancial} services, particularly stock market trading, constitute a pivotal element of financial systems globally, with an estimated capitalization exceeding 89 trillion U.S. dollars in 2020 \cite{chodorow2021stock}. Concurrently, advancements in Internet technology have profoundly transformed traditional stock market mechanisms, ushering in an era of electronic markets \cite{chovancova2018changes}. This evolution has facilitated automated order execution, culminating in reduced transaction fees, enhanced market efficiency, and augmented information accessibility and transparency for investors. Furthermore, the quest for high returns through investment has increasingly focused attention on stock trend forecasting in recent years, marking it as a highly anticipated and prevalent investment approach. Specifically, stock trend forecasting employs a blend of manual and technical methodologies to evaluate and anticipate stock market directions \cite{li2019individualized}, utilizing tools such as Moving Average Convergence Divergence (MACD), Stochastic Oscillator, and Exponential Moving Average (EMA).
However, the real-world scenario presents a challenge due to the inherent unpredictability and pervasive noise in stock information, characterized by stochastic market dynamics, non-linear behavior in large-scale markets with billions in capital and millions of participants, information asymmetries, and the erratic conduct of traders. These factors collectively impede the precision of trend predictions \cite{chen2011stock}. Consequently, it becomes imperative to incorporate an analysis of the aforementioned implicit and explicit biases, referred to as exformation, to enhance the accuracy of stock trend forecasts.

Recently, Machine Learning (ML) applications in various industries have shown their potential and return in stock trend forecasting, in particular, \cite{jiang2021applications, CHEN2021853}. To optimize forecast performance, it is a common practice to mine fundamental and technical factors in the temporal information of historical price patterns by feeding relative features into models to analyze the market movements. Typically, models such as Transformers \cite{ding2020hierarchical}, Long Short-Term Memory (LSTM) \cite{hochreiter1997long}, and Gated Recurrent Unit (GRU) \cite{chung2014empirical, gupta2022stocknet} use features based on price, relationship, and sequential information to capture non-linear patterns in multivariate time series. To further exploit the potential interdependencies between different factors, a range of approaches have been proposed. For example, State Frequency Memory (SFM) \cite{zhang2017stock} uses the discrete Fourier transform to capture multi-frequency trading patterns; Hierarchical Concept-Oriented Shared Information Mining for stock trend forecasting (HIST) \cite{xu2021hist}, and Graph Attention Networks (GATs) \cite{casanova2018graph} regard variables as nodes to construct graphs to mine the interactions between regard variables within the same time stamp.

Although previous research has demonstrated impressive performance on this task, two key challenges remain unresolved. Firstly, discrepancies between the noise distribution in training datasets and that in actual-world data present a formidable obstacle for models grounded in empirical knowledge. Such an approach is prone to overfitting and exhibits pronounced inductive biases. This issue is accentuated by the foundational premise of empirical design, which posits that underlying principles are universally applicable across diverse contexts. However, models are typically constructed through a fixed methodology based on expert insights, thereby impeding their adaptability to the evolving noise distribution in stock movements, consequently curtailing the model's genuine profitability. Secondly, market frictions engender transient limitations in the market, culminating in a price pressure effect. The equilibrium point and immediate returns attributed to each price pressure incident offer insights into both the short-term and long-term implications of these constraints. When combined with the noise inherent in raw stock data, they contribute to the multi-level noise observed in pricing. Consequently, the models disproportionately accentuate mixed noise patterns while simultaneously overlooking and misinterpreting authentic profit signal patterns.

\begin{figure}[!t]
    \centering
    \includegraphics[width=\linewidth]{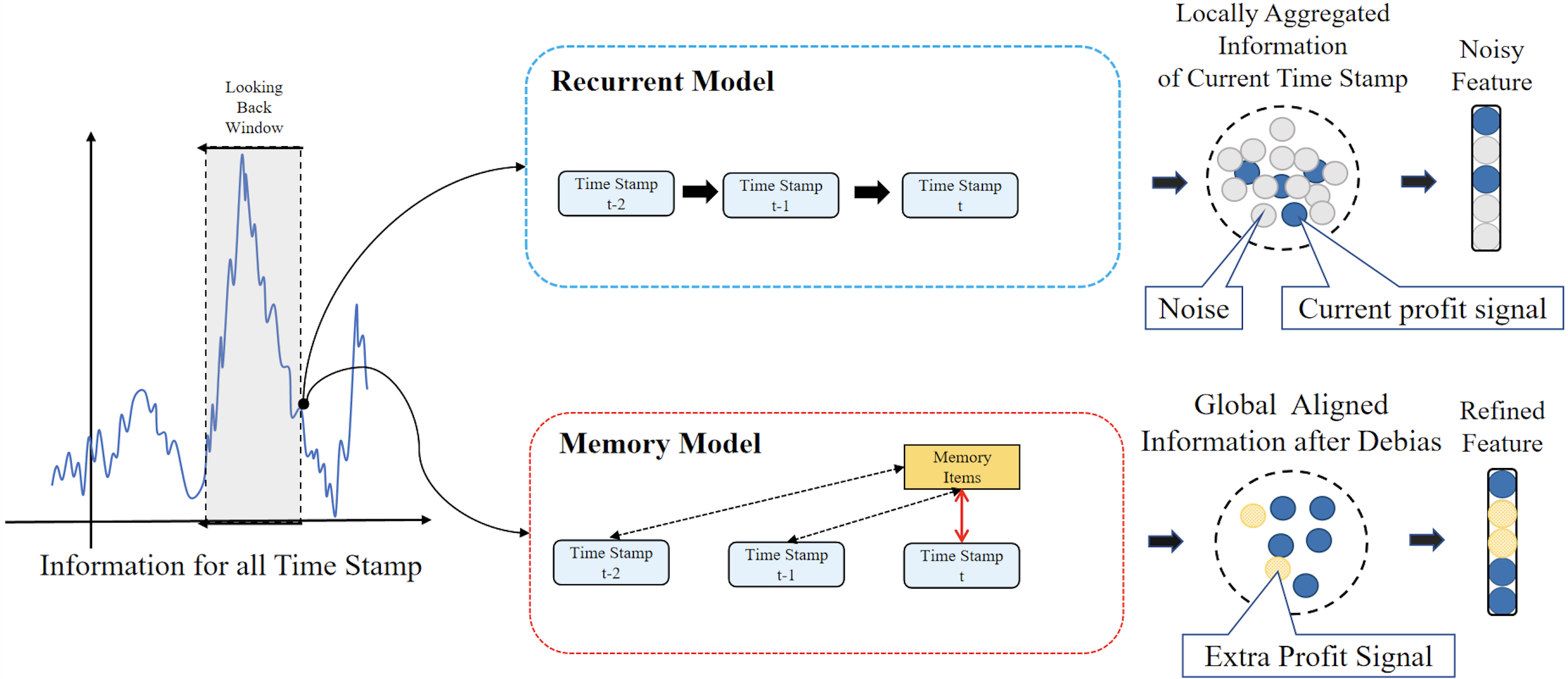}
    \caption{Illustration of recurrent model and memory model in the real stock market. The recurrent model only depends on the state at $t-1$, and it will tend to learn noise patterns from historical data. However, the memory spans the global information flow with multiple previous timestamps, which helps capture extra profit signals for the current timestamp. In addition, the local aggregation will fit the noise in the current data. The memory model uses global alignment and the knowledge collected in the past to achieve denoising.}
    \label{fig:10}
\end{figure}

The aforementioned challenges coexist and interact within a volatile market environment, exacerbating the risks associated with model forecasting. It is crucial to efficiently capture profit signals and succinctly articulate their generalized attributes. Regrettably, these signals are often sparse and indistinct, eluding detection at individual timestamps on a local scale, a common limitation in existing methodologies \cite{hochreiter1997long, chung2014empirical}. For instance, recurrent-based methods exhibit limited proficiency in capturing multi-scale temporal coherence. We propose that by filtering out chaotic noise—for instance, isolating the profit signal from a noisy dataset—such temporal coherence could be enhanced, facilitating the recognition of events of varying durations. Nevertheless, as depicted in Fig.~\ref{fig:10}, both LSTM and GRU models tend to diminish long-range temporal consistency within a Markov chain, where the state at time $t$ is solely influenced by the state at time $t-1$.

In contrast, by considering the global information flow, profit signals could form a slender yet significant thread across disparate timestamps. Therefore, our approach focuses on retaining essential information throughout various training phases to trace profit signal patterns. As illustrated in the lower section of Fig.~\ref{fig:10}, our method involves the prior state capturing a general, representative pattern in memory, serving as a reference in processing the subsequent $t+1$ state, thus maintaining temporal consistency. This strategy contrasts with an `exhausted LSTM' approach, which links the current state to all preceding individual states in the memory bank. Our method boasts two primary advantages: firstly, our memory module learns to represent prolonged events, for example, an event spanning from $t$ to $t+10$, as a singular, consolidated representation stored in memory, rather than separately recording all ten features of each state. This reduces complexity from $\mathcal{O}(T^2)$ to $\mathcal{O}(T)$, where $T$ represents all timestamps. Secondly, our approach fosters event consistency by deliberately suppressing the activation of irrelevant events and noise. Consequently, the summarized pattern in memory emerges as more distinctive and meaningful than disorganized noise.

In this paper, we introduce a novel Multi-Scale Temporal Memory and Debiasing Framework (MTMD) for forecasting stock trends. Our approach integrates a pioneering memory module to ensure temporal consistency by transferring aggregated long-range profit features from the preceding timestamp to the current one, as illustrated in Fig. \ref{fig:10}. Furthermore, to mitigate overwhelming complexity, we employ a learnable embedding paired with external attention, wherein each element of the embedding represents an individual, event-level profit signal. This strategy significantly reduces the volume of long-range temporal information. To further bolster temporal coherence, we implement a negative inhibition mechanism that suppresses irrelevant noise and events during the activation computation with prior memory items. This approach is independent of time intervals and clusters the enduring timestamps of an event through self-similarity, a critical feature for linking distant timestamps in recurrent models.

Subsequent to the development of MTMD, we conducted comprehensive experiments on real-world stock market datasets, benchmarking against nine existing models. The results demonstrate that MTMD surpasses these baselines across multiple evaluation metrics. Notably, our method exhibits an improvement of up to 6.0\% and 4.3\% in Information Coefficient (IC) \cite{lin2021learning} and Rank IC \cite{li2019individualized} scores, respectively, with IC representing the Pearson correlation coefficient and Rank IC denoting Spearman’s rank correlation coefficient between actual and predicted change rates. Further analyses were conducted to assess the individual components of MTMD, reinforcing its reliability. In summary, the primary contributions of our work are as follows:
\begin{itemize}
\item We introduce a groundbreaking framework designed to capture and chronicle patterns of actual profit signals concealed within the information flow using memory items, thereby attenuating the effects of empirical design inconsistency and pronounced model bias.
\item Our framework processes features across multiple scales, enabling an exhaustive extraction of local information and differentiation at both global and local scales. This facilitates a more nuanced understanding of stock features in response to event perturbations at varying intensities.
\item Through extensive experimentation on real stock market data, we establish new benchmarks for state-of-the-art (SOTA) performance, demonstrating improvements of 6\% and 4\% in IC and Rank IC metrics compared to prior SOTA models. Additionally, the efficacy of the MTMD framework is robustly validated.
\end{itemize}
The remainder of the paper is organized as follows. Section \ref{Related Works} introduces background concepts and related work on Stock Trend Forecasting. Section \ref{Methodology} describes the proposed framework. Section \ref{Exp} presents experiments and discusses our findings. Finally, Section \ref{conclusion} concludes the paper and provides directions for future work.

\section{Related work}\label{Related Works}
In this section, we review the relevant literature streams regarding traditional financial research, ML, and deep learning to position our research findings from extant research. 

In traditional financial research, the technology analysis commonly used in stock trading in the early days was mainly constructed based on mathematical logic and rules of thumb. On the one hand, mathematical logic is to directly predict prices based on historical prices. common indicators such as MA (moving average) and MACD (moving average convergence and divergence) are based on historical intraday prices (high, low, open, close) for a certain length of time, whose difference calculation reflects the price trend in the future period. On the other hand, the core of the rule of thumb is supply and demand, which means judging the behavior of going long and short by price and volume, indirectly predicting the price at the next moment, For example, RSI (relative strength index) determines the strength of the buying and selling forces of the long and short sides in the market that by comparing the rise and fall of a single stock price even the whole market index in a period of time, so as to judge the future market trend. However, traditional quantitative financial methods are mainly limited to subjective experience, whose assumptions and inferences of those trend indicators are too ideal making it difficult to capture the signals sent by high-dimensional data \cite{gavrishchaka2003volatility}. 

Since its development nearly 20 years ago, the use of artificial intelligence technology in stock trend forecasting has significantly increased traders' decision-making effectiveness and helped them comprehend transaction timing more precisely. Current approaches frequently make use of ML and deep learning techniques \cite{kumbure2022machine}. Typically ML is widely used with technical analysis, which is a method to mine factors such as price, temporal, volatility, chart, and statistics. In the early stage of ML development, most researchers focus on stationary regression models. For example, the Autoregressive Model (AR) \cite{li2016stock} and Autoregressive Integrated Moving Average model (ARIMA) \cite{ariyo2014stock} have been widely applied for a long time but they are limited because of the nonlinearity of stock price. In addition, models built upon extremely versatile tree models are also popular, such as decision tree \cite{wu2006effective, kim2018forecasting, jin2020stock}, random forest \cite{ballings2015evaluating}, gradient boosting decision tree \cite{le2019fast}. Apart from that, Artificial Neural Networks (ANN) \cite{kim2000genetic, khashei2010artificial, kara2011predicting}, Genetic algorithms (GA) \cite{kim2000genetic, cheng2010hybrid, aguilar2015genetic}, and the Hidden Markov Model (HMM) \cite{xiaoning2019stock} are three methods used to forecast the behavior of the financial markets. Moreover, Support Vector Machines (SVMs) have shown better performance in forecasting stock market changes \cite{lin2013svm}. A multi-task recurrent neural network with high-order Markov Random Fields (MRFs) is introduced to forecast the direction of stock price movement \cite{li2019multi}. Owing to their excellent efficiency, interpretability, and controllability, these methodologies and models continue to play a significant role in the financial industry today. However, the upper limits of traditional ML and statistical factor analysis depend on expert knowledge, and when expert knowledge is biased and inadequate, the models do not work as expected in real markets. 

However, with the deepening of research, we find that traditional ML is not enough to completely mine representative features from financial data \cite{zhang2022intraday}. Particularly, following the rapid growth of informatization, technical analysis will be subject to complex and unidentified market interference information, making the predicted results unsatisfactory \cite{picasso2019technical}. To bridge the gap in predictive power between the real market and expert knowledge, an increasing number of studies use deep learning models to extract dense vectorized features of financial data and avoid shaking caused by noise \cite{wu2018hybrid, jiang2021applications}. For instance, models like recurrent neural networks (RNN) \cite{zhang2018new}, LSTM \cite{hochreiter1997long, chen2015lstm, nelson2017stock} and GRU network \cite{chung2014empirical, xu2022stock} have been deployed in the market forecast and have achieved competitive baseline results. Recently, a novel model of SFM recurrent network is proposed to discover multi-frequency trading patterns for stock price volatility forecast \cite{zhang2017stock}. Feng \textit{et al.} proposed an ALSTM framework, which combines the attention mechanism with a dual-stage recurrent neural network \cite{feng2018enhancing}, proving that the attention mechanism is conducive for time series forecasts. Following this, the Transformer framework is applied to the dynamic forecast of stocks in Ding \textit{et al.}'s study, where the success of adopting unconstrained time step length showed that the attention mechanism is also effective for mining global temporal information \cite{ding2020hierarchical}. Multiscale information in financial time series has been extensively studied by \cite{dacorogna1996changing}. Geva and Amir decompose time series into different scales through wavelet transform and extract features at each scale through different neural networks to obtain predictions \cite{geva1998scalenet}. Fern \textit{et al.} applied extreme learning machine (ELM) and discrete wavelet transform (DWT) to capture scale properties \cite{fernandez2019meta}. Liu \textit{et al.} aimed to automatically learn multi-scale information in stock data, and proposed a multi-scale bidirectional deep neural network Network (MTDNN) \cite{liu2020multi}. Chen \textit{et al.} established temporal relationships for node embedding through temporal convolutional networks to achieve multivariate time series forecasting \cite{chen2022multi}.

Besides, the cutting-edge research in the financial field is based on the framework of deep learning, Graph Neural Networks (GNNs)-based methods are proposed to establish the correlation between the stock concepts and industry sector information for returns \cite{xu2022hgnn}. For example, Graph Convolutional Networks (GCNs) are used to represent the relationships between shareholders and the industry \cite{chen2018incorporating, feng2019temporal}. In order to predict the future direction of the stock market,  present a hierarchical attention network for stock prediction (HATS) that makes use of relational data \cite{kim2019hats}. GATs leverage masked self-attention layers on graph-structured data in stacked layers. The nodes have different weights so they can attend over their neighborhoods’ features without requiring any kind of costly matrix operation (such as inversion) or depending on knowing the graph structure upfront \cite{casanova2018graph}. HATS uses relational data to forecast stock markets. With an emphasis on the dynamic correlation between stocks and specified ideas, hierarchical excavation of more comprehensive information achieves a more precise trend forecast for stocks \cite{kim2019hats}.

\begin{table}[!t]
\caption{Variable Symbols and Definitions}
\label{v1}
\resizebox{\linewidth}{!}{
\begin{tabular}{l@{}ll@{}}
\hline
&Variable & Definition \\
\hline
&$X^t$ & Raw stock features on date $t$\\
&$N_s$ & Number of target stocks\\
&$N_c$ & Number of predefined concepts\\
&$t$ & Time steps\\ 
&$\hat{\vec{p}}^t$ & Predicted price changes\\
&$p_i^t$ & Real change rate\\
&$\theta$ & Different modules\\
&$k$ & Number of memory items\\ 
&$\vec{h}^{t,\theta}_i$ & Stock i's feature vector\\
&$\hat{\vec{h}}^{t,\theta}_i$ & Estimated feature vector of module prediction\\
&$\vec{y}^{t,\theta}_i$ & Module input or output vector\\
&$\vec{q}^{\theta}_k$ & Feature vector of each memory item\\
&$U^{t,\theta}_k$  & The set of indices of each memory item\\
&$\vec{m}^{\theta}_k$ & Memory vector of each memory item\\
\hline
\end{tabular}}
\end{table}

\section{Methodology}\label{Methodology}
In our research, we introduce the MTMD framework, where we utilize the HIST \cite{xu2021hist} as its backbone to demonstrate MTMD's effectiveness. Furthermore, to showcase MTMD's plug-and-play capability and universality, we conduct a series of experiments using an MLP as the backbone in Experiment \ref{ob}. This highlights MTMD's adaptability and superior performance in enhancing stock trend predictions across different models.

The raw stock features of $N_s$ target stocks on date $t$ can be denoted by $X^t=\{\vec{x}_1^t, \vec{x}_2^t,\cdots, \vec{x}_{N_s}^t\}$ with $N_c$ predefined concepts $C^t=\{\vec{c}_1^t, \vec{c}_2^t,\cdots, \vec{c}_{N_c}^t\}$. Each raw stock feature $\vec{x}_i^t$ of stock $i$ on date $t$ is a vector, which contains a collection of features to stock $i$ on date $t$, including opening price, closing price, volume, \textit{etc}. Here, we need to predict the the stock trend $\hat{\vec{p}}^t=\{\hat{p}_1^t, \hat{p}_2^t, \cdots, \hat{p}_{N_s}^t\}$ of the input stocks. Therefore, considering a stock $i$ on date $t$, our problem can be formulated as a mapping $f$ (a deep model in our approach) with parameters $\vec{W}$ such that
\begin{equation}
    \hat{p}_i^t=f(\vec{x}_i^t, C^t;\vec{W}),
\end{equation}
where $\hat{p}_i^t$ is the predicted trend (change rate) and its corresponding ground truth is defined as $p_i^t=\frac{price^{t+1}-price^{t}}{price^{t}}$, which is the real change rate of stock $i$ from date $t$ to $t+1$. Here, we hope that the predicted trend $\hat{p}_i^t$ is as close to the real trend $p_i^t$ as possible. The frequently used notations are summarized in Table~\ref{v1}.

\begin{figure}[!t]
    \centering
    \includegraphics[width=\linewidth]{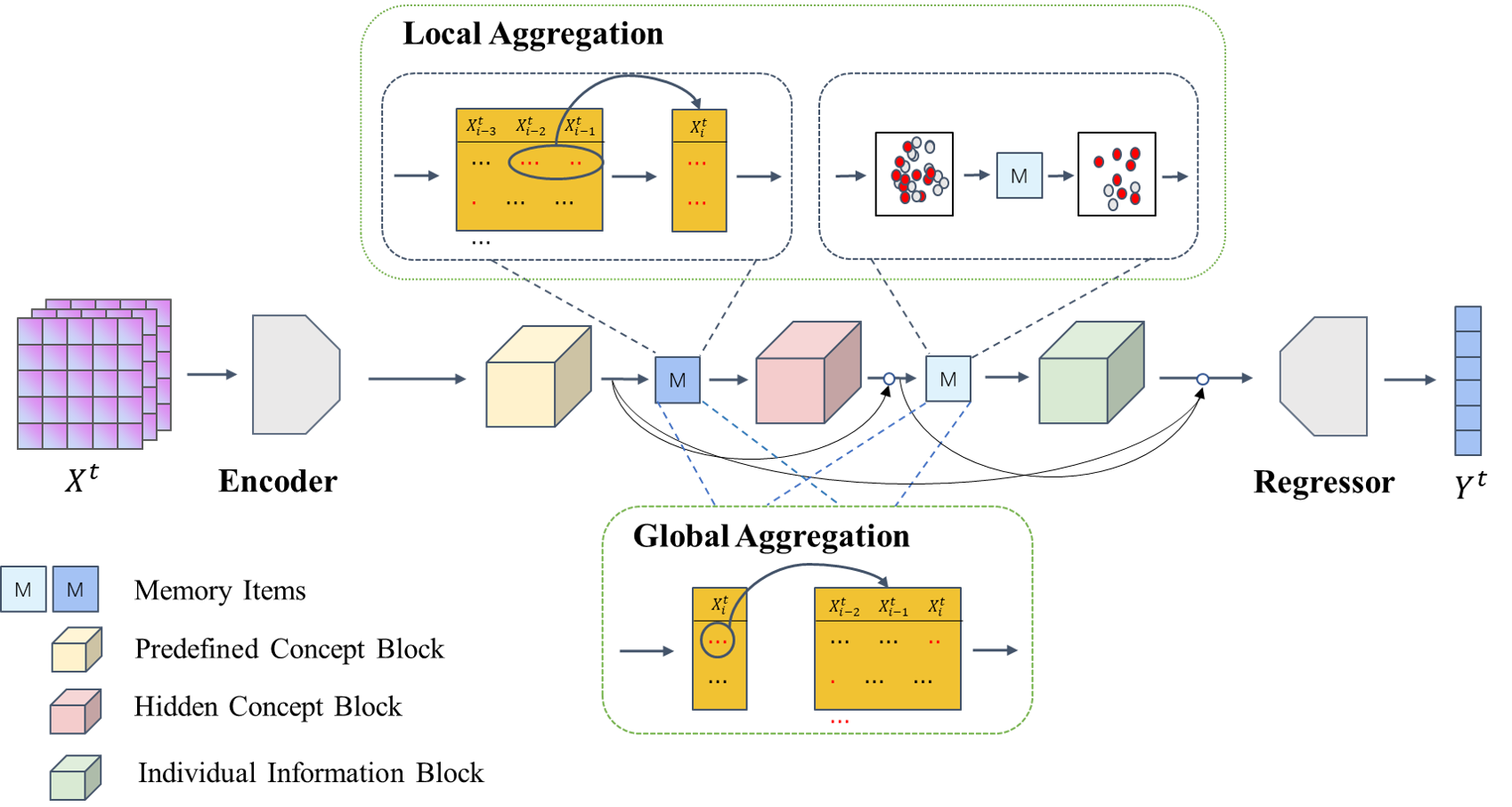}
    \caption{Illustrating the pipeline of MTMD. The global and local aggregations modules denote capturing and learning multi-scale temporal pieces of information and debiased.}
    \label{fig:overview}
\end{figure}

In our approach, we first define a \textit{stock-concept feature extractor} as the starting point because this is the fundamental component of our MTMD framework. Then, by introducing memory items into the stock-concept feature extractor, we formulate a \textit{memory-assisted stock-concept feature extractor} to extract deep features from the raw stock feature $X^t$. Finally, we can get the prediction $\hat{\vec{p}}^t$ by combining the memory-assisted stock-concept feature extractor and \textit{stock trend regressor}. The \textit{stock trend regressor} is a predictor based on the output features from the memory-assisted stock-concept feature extractor. Thus, our MTMD framework consists of the \textit{memory-assisted stock-concept feature extractor} and \textit{stock trend regressor}. Whether it is the \textit{stock-concept feature extractor} or the \textit{memory-assisted stock-concept feature extractor}, they all have three modules: Predefined concept module, hidden concept module, and individual information module. The basic architecture is shown in Fig. \ref{fig:overview}.

\begin{figure}[!t]
    \centering
    \includegraphics[width=\linewidth]{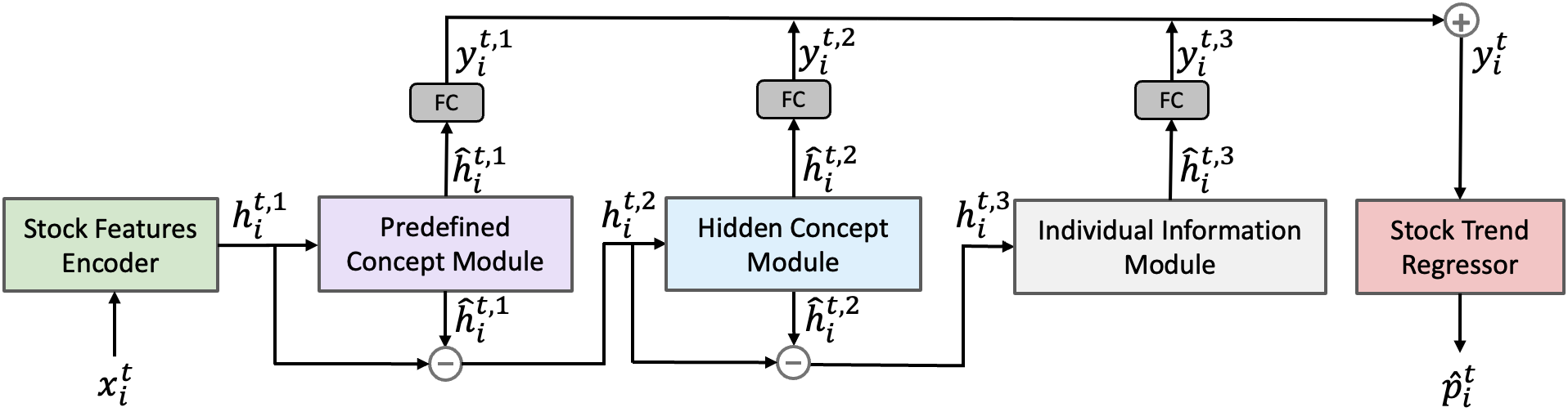}
    \caption{The framework of stock-concept feature extractor.}
    \label{fig:frame1}
\end{figure}

\begin{figure}[!t]
    \centering
    \includegraphics[width=\linewidth]{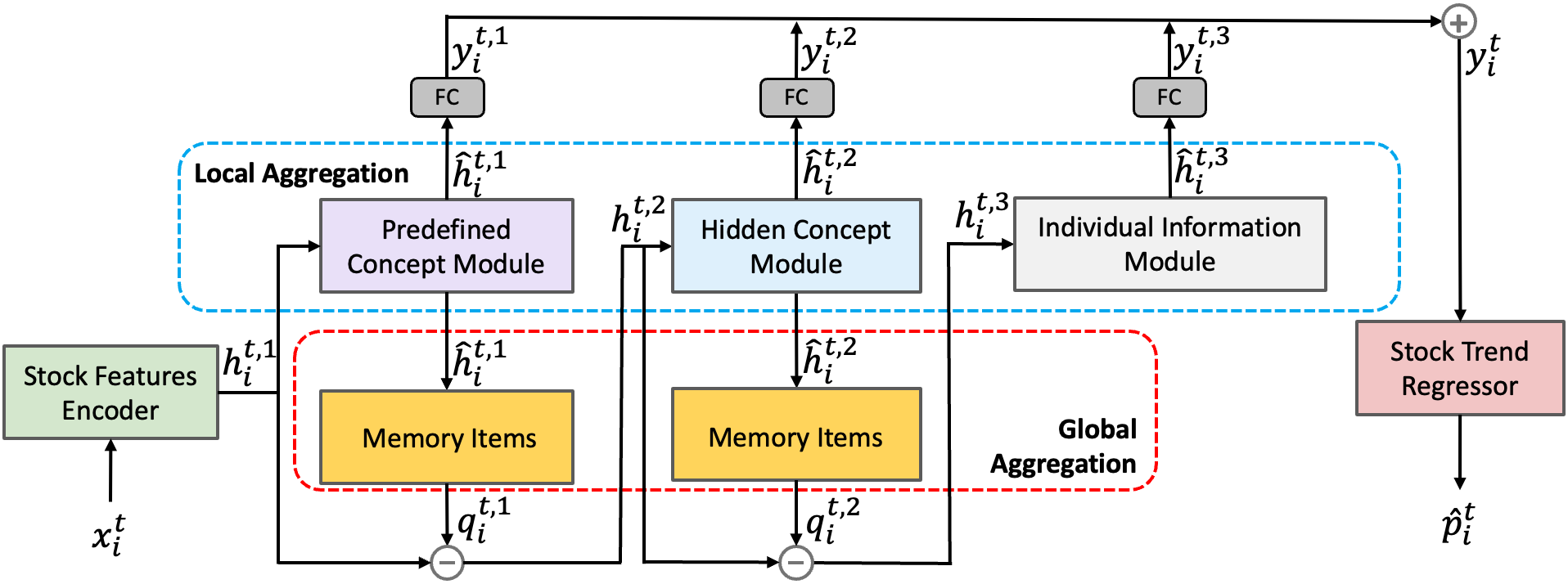}
    \caption{The framework of memory-assisted stock-concept feature extractor.}
    \label{fig:frame2}
\end{figure}

\begin{itemize}
    \item \textbf{Stock-concept feature extractor:} It consists of a stock feature encoder and three modules shown in Fig. \ref{fig:frame1}: predefined concept module, hidden concept module, and individual information module, which are denoted by $\theta=1,2,3$ respectively. When considering the date $t$ and stock $\vec{x}_i^t$, the encoded stock features provide initial temporal embedding $\vec{h}^{t,1}_i \in \mathbb{R}^{L}$ with $\theta=1$ by using historical price data, and then the other three modules further extract various supplementary information using related stock concepts. The modules are organized through doubly residual architecture. For each module $\theta$, we input $\vec{h}^{t,\theta}_i$ obtained from the previous module $\theta-1$ and output the stock-concept feature $\hat{\vec{h}}^{t,\theta}_i \in \mathbb{R}^L$. This stock-concept feature $\hat{\vec{h}}^{t,\theta}_i$ contains the effect of its corresponding module, which will be removed from the current input feature to generate the input of the next module. Thus, we have $\vec{h}^{t,\theta+1}_i=\vec{h}^{t,\theta}_i-\hat{\vec{h}}^{t,\theta}_i$. Besides, it can formulate a forecast feature $\vec{y}^{t,\theta}_i \in \mathbb{R}^L$ by using a fully connected (FC) network. Finally, we get the eventually extracted feature $\vec{y}^t_i=\vec{y}^{t,1}_i+\vec{y}^{t,2}_i+\vec{y}^{t,3}_i$.
    
    \item \textbf{Memory-Aware Feature Aggregators:}
    This framework is shown in Fig. \ref{fig:frame2}. We add a memory block for the predefined concept module and the hidden concept module, where the memory block is composed of $K$ memory items. These memory blocks are global and collect all information in previous timestamps. The predefined concept module, hidden concept module, and individual information module are trained only in the current timestamp, thus they are local. Here, we can say that memory blocks are global aggregation and other modules are local aggregation. Shown as the blue dotted box in Fig. \ref{fig:frame2}, it is the local aggregation, which uses a stock-concept feature extractor for local encoding. Shown as the red dotted box in Fig. \ref{fig:frame2}, it is the global aggregation. For the first two modules $\theta\in\{1,2\}$, predefined concept module or hidden concepts module, we input the stock-concept feature $\hat{\vec{h}}^{t,\theta}_i$ to its memory block, and get a low-noise and global feature $\vec{q}^{t,\theta}_i$. This global feature $\vec{q}^{t,\theta}_i$ will be removed from the current input feature to generate the input of the next module. Thus, we have $\vec{h}^{t,\theta+1}_i=\vec{h}^{t,\theta}_i-\vec{q}^{t,\theta}_i$.
    \item \textbf{Stock trend regressor} is a feed-forward network, which finally sums the different types of features given by the feature extractor above to make forecasts.
\end{itemize}

\subsection{Stock-concept Feature Extractor}
\subsubsection{Stock Feature Encoder} 
Given the matrix ${X}^t=\{\vec{x}_1^t, \vec{x}_2^t, \cdots, \vec{x}_{N_s}^t\} \in \mathbb{R}^{N_s \times L'}$, the raw stock features of $N_s$ target stocks, with each of the row vector $\vec{x}_i^t \in \mathbb{R}^{L'}$ contains $K$ elements composed of historical trading prices and volumes of stock $i$ within a looking back window, the stock feature encoder extracts temporal information in them and compress the information as a denser embedding matrix in a lower dimension. We apply an encoder to generate the initial embedding:
\begin{equation}
    \vec{h}^{t,1}_i = {\rm Encoder}(\vec{x}^t_i) \label{eq311}
\end{equation}
where $\vec{h}^{t,1}_i \in \mathbb{R}^{L}, L<L'$ represents temporal information in stock $i$, and each of them is the row of matrix $\vec{H}^{t,1} \in \mathbb{R}^{N_s \times L}$.

\subsubsection{Predefined Concept Block}
The predefined concepts are manually defined by a category of stocks, such as different stock market sectors. For example, Google and Facebook are classified as communication services, while Walmart and Coca-Cola are classified as daily consumer goods. These concepts connect companies and build a bipartite graph of stock concepts, from which meaningful shared information can be mined to obtain more accurate forecasts.

Therefore, the predefined concept module is designed to learn such a dynamic concept embedding from the graph built on each date. Given $N_c$ predefined concepts $C^t$ on date $t$ and temporal embedding $\vec{h}^{t,1}$, we initialize the embedding of each concept $\vec{c}^t_j$, denoted by $\vec{e}^{t,1}_j$, as the weighted average over the encoded features of stocks connected with the concept:
\begin{equation}\label{eq312}
    \vec{e}^{t,1}_j=\sum\nolimits_{i \in D^{t,1}_j}\alpha^t_{i,j}\vec{h}^{t,1}_i,
\end{equation}
where $\vec{e}^{t,1}_j \in \mathbb{R}^{L}$ is the initial embedding of concept $\vec{c}^t_j$ on date $t$, $D^{t,1}_j$ is the set of stocks connected with concept $\vec{c}^t_j$, and $\alpha^t_{i,j}$ is the weight indicating the degree of contribution made by stock $i$ to concept $\vec{c}^t_j$. To measure this, we take advantage of the market capitalization $\delta^t_{i}$ of stock $i$ on date $t$ and compute its proportion over all connected stocks as
\begin{equation}\label{eq313}
    \alpha^t_{i,j}=\frac{\delta^t_i}{\sum_{i' \in D^{t,1}_j}\delta ^t_{i'}}.
\end{equation}

The above initial embedding can be learned from a partially connected graph. However, some of the significant links between stocks and predefined concepts may not be covered by the input data, while some of the links may be incorrectly connected. Thus, we correct the concept embedding by using a fully connected graph with soft links instead.

To model the strength of these soft links, we compute the cosine similarity between each pair of stock feature $\vec{h}^{t,1}_i$ and concept embedding $\vec{e}^{t,1}_j$, and then normalize these scores as probability using the softmax function:
\begin{equation}\label{eq314a}
\beta_{i,j}^{t,1}=\cos\left(\vec{h}^{t,1}_i,\vec{e}^{t,1}_j\right)=\frac{\vec{h}^{t,1}_i \cdot \vec{e}^{t,1}_j}{||\vec{h}^{t,1}_i|| \cdot ||\vec{e}^{t,1}_j||},
\end{equation}
\begin{equation}\label{eq314b}
    \hat{\alpha}^{t}_{i,j}=\frac{\exp(\beta_{i,j}^{t,1})}{\sum^{N_s}_{i'=1}\exp(\beta_{i',j}^{t,1})},
\end{equation}
where $\cos(\vec{a}, \vec{b})$ denotes cosine similarity between two vectors. The embedding of the concept will be updated using these normalized probabilities and a linear layer with ${\rm LeakyReLU}$ activation as:
\begin{equation}\label{eq315}
    \vec{e}^{t,1}_j={\rm LeakyReLU}\left[\vec{W}'_1\left(\sum\nolimits^{N_s}_{i=1} \hat{\alpha}^{t}_{i,j}\vec{h}^{t,1}_{i}\right)+\vec{b}'_1\right],
\end{equation}
where $\vec{e}^{t,1}_j \in \mathbb{R}^L$ is the predefined concept embedding after correction, and $\vec{W}'_1$ and $\vec{b}'_1$ are learnable parameters. 
In order to mine the relationship between concept embedding and stock features more effectively, the concept embedding will be aggregated into the stock features by the memory-aware feature aggregator (see \ref{memory}), obtaining a stock-concept feature $\hat{\vec{h}}^{t,1}_i$ with a forecast feature $\vec{y}^{t,1}_i$ for each stock $i$ on date $t$.

\subsubsection{Hidden Concept Block}
Predefined concepts can effectively capture shared information in target stocks. However, the scope of predefined stocks may be limited, and some latent correlations also influence the stock movements. Therefore, we try to mine hidden concepts beyond predefined ones and obtain the corresponding embedding which is shown as the blue block in Fig. \ref{fig:frame2}.

Based on doubly residual architecture, we denote by the input of hidden concept Module $\vec{h}^{t,2}_i$, which can be defined as
\begin{equation}
    \vec{h}^{t,2}_i=\vec{h}^{t,1}_i-\vec{q}^{t,1}_i,
\end{equation}
where $\vec{h}^{t,2}_i \in \mathbb{R}^L$ represents the information of stock $i$ not captured by the previous module.

To mine the hidden concepts, we first assume that there are hidden concepts on date $t$, each of their embedding $\vec{e}^{t,2}_j$ is initialized as $\vec{e}^{t,1}_j$. Then we compute the similarity scores ${\ensuremath{\beta}}^t_{i,j}$ between each of the input features and initial hidden concept embedding as:
\begin{equation}
\beta_{i,j}^{t,2}=\cos\left(\vec{h}^{t,2}_i,\vec{e}^{t,2}_j\right)=\frac{\vec{h}^{t,2}_i \cdot \vec{e}^{t,2}_j}{||\vec{h}^{t,2}_i|| \cdot ||\vec{e}^{t,2}_j||}.
\end{equation}

In order to dig deeper into conceptual relationships, we build a relationship between each stock $i$ and the concept with the greatest similarity to this stock, that is $j^*=\arg\max_j\{\beta^{t,2}_{i,j}\}$, and get a new stock set $D^{t,2}_j$ for each concept $\vec{c}^t_j$. To prune this bipartite graph and make it more informative, we remove the relationships that are in $D^{t,1}_j$ from $D^{t,2}_j$.


Thus, we compute the hidden concept embedding accordingly:
\begin{equation}\label{eq315}
    \vec{e}^{t,2}_j={\rm LeakyReLU}\left[\vec{W'}_{2}\left(\sum\nolimits_{i \in D^{t,2}_j}\beta^{t,2}_{i,j}\vec{h}^{t,2}_{i}\right)+\vec{b}'_{2}\right],
\end{equation}
where $\vec{e}^{t,2}_j \in \mathbb{R}^L$ is the embedding of hidden concept after correction, and $\vec{W}'_2$ and $\vec{b}'_2$ are learnable parameters. Similar to the predefined concept module, the embedding obtained from hidden concept block will also be aggregated into the stock features by the memory-aware feature aggregator, obtaining a stock-concept feature $\hat{\vec{h}}^{t,2}_i$ with a forecast feature $\vec{y}^{t,2}_i$ for each stock $i$ on date $t$.

\subsubsection{Individual Information Block}
The previous two modules capture shared information between the target stocks, but it remains some other features possessed by each particular stock individually. Therefore, like the previous module, we extract the individual feature as the residual remained by the last module's input. Shown as the grey block in Fig. \ref{fig:frame2}, we have
\begin{equation}
    \vec{h}^{t,3}_i=\vec{h}^{t,2}_i-\vec{q}^{t,2}_i,
\end{equation}
where $\vec{h}^{t,3}_i \in \mathbb{R}^{L}$ represents the individual information of stock $i$. In this module, it does not need to be integrated with the memory-aware feature aggregator, which directly obtains a stock-concept feature $\hat{\vec{h}}^{t,3}_i$ with a forecast feature $\vec{y}^{t,3}_i$ for each stock $i$ on date $t$. Thus, we have
\begin{equation}\label{eq401}
    \hat{\vec{h}}^{t,3}_i={\rm LeakyReLU}\left[\vec{W}_3\vec{h}^{t,3}_i+\vec{b}_3\right],
\end{equation}
where $\vec{W}_3$ and $\vec{b}_3$ are learneable FC parameters.

\subsection{Memory-aware Feature Aggregator} \label{memory}
As mentioned in the previous section, two memory-aware feature aggregators are integrated into the predefined and hidden concept module. Since aggregators in these two modules are similar, we unify their working process as one pipeline for convenience during the later explanation, where $\theta$ is used as the identifier of each module ($\theta=1$ for predefined concept module and $\theta=2$ for hidden concept module). Overall, the memory-aware feature aggregators perform multi-scale feature aggregation with the help of memory items $\vec{M}^{\theta}\in\mathbb{R}^{K\times L}$ that can store historical stock-concept patterns. Each of the items is a vector of identical size with the concept features $\vec{e}^{t,\theta}_i\in\mathbb{R}^L$ that will be aggregated, containing concept information on date $t$ extracted by module $\theta$. Two operations, aggregating and memorizing, are performed by the aggregators, shown in Fig. \ref{fig:frame2}, and details about them will be discussed as follows. 

\subsubsection{Aggregation}
Aggregation operation carries out multi-scale feature aggregation. On the local scale, the aggregator fuses the information in the concept features and stock features on date $t$ together. On the global scale, the memory block of the aggregator combines the local features with profit signal patterns recorded by the memory items across the whole time flow included in the training data.

\textbf{Local aggregation.} 
Shown as the red dotted box in Fig. \ref{fig:frame2}, given all input stock feature $\vec{h}^{t,\theta}_i$ and concept embedding $\vec{e}^{t,\theta}_j$, we compute the similarity between each of the stock-concept pairs and apply the softmax function to normalize these scores as probability along the dimension of all the concepts. Thus, we have:
\begin{equation}\label{eq325}
    \beta_{i,j}^{t,\theta}=\cos\left(\vec{h}^{t,\theta}_i,\vec{e}^{t,\theta}_j\right)=\frac{\vec{h}^{t,\theta}_i \cdot \vec{e}^{t,\theta}_j}{||\vec{h}^{t,\theta}_i|| \cdot ||\vec{e}^{t,\theta}_j||},
\end{equation}
\begin{equation}\label{eq326}
    \gamma^{t,\theta}_{i,j}=\frac{\exp(\beta_{i,j}^{t,\theta})}{\sum_{j' \in D^{t,\theta}_i}\exp(\beta_{i,j'}^{t,\theta})},
\end{equation}
where $D^{t,\theta}_i$ is the set of concepts that connect with stock $i$ and $\gamma^t_{i,j}$ represents the importance of a concept to the stock it connects with, which will be used to aggregate the concept features as:
\begin{equation}\label{eq327}
    \hat{\vec{h}}^{t,\theta}_i={\rm LeakyReLU}\left[\vec{W}_\theta\left(\sum\nolimits_{j \in D^{t,\theta}_i}\gamma^{t,\theta}_{i,j}\vec{e}^{t,\theta}_{j}\right)+\vec{b}_\theta\right],
\end{equation}
where $\hat{\vec{h}}^{t,\theta}_i$ is the stock-concept feature on date $t$ corresponding to stock $i$, and $\vec{W}_\theta$ and $\vec{b}_\theta$ are learnable parameters. It is computed as the weighted sum of concept features that are connected to stock $i$.

\textbf{Global aggregation.}
Shown as the blue dotted box in Fig. \ref{fig:frame2}, the global aggregation process aggregates the profit signal patterns recorded in the memory items. The locally aggregated features are used as queries to retrieve related profit signals from the memory items, obtaining globally refined features. The similarity scores between each query $\hat{\vec{h}}^{t,\theta}_i \in \mathbb{R}^{L}$ and memory item $\vec{M}^{\theta}\in \mathbb{R}^{K\times L}$ are computed by matrix product, which can formulate a correlation vector, denoted by $\vec{b}_i^{t,\theta} \in \mathbb{R}^{K}$. Thus, we have
\begin{equation}
    \vec{b}_i^{t,\theta}=\vec{M}^{\theta}\hat{\vec{h}}^{t,\theta}_i.
\end{equation}

For our problem, there are $N_s$ stocks in a batch. In order to avoid excessive variance, we adopt the technique of batch normalization among these $N_s$ stocks. Here, we denoted by $(\vec{b}^{t,\theta}_i)_k$ the $k$-th element of vector $\vec{b}^{t,\theta}_i$. After the batch normalization, we can change $\vec{b}^{t,\theta}_i$ to $\vec{v}^{t,\theta}_i$ by using
\begin{equation}\label{eq321}
    (\vec{v}^{t,\theta}_{i})_k=\frac{\exp((\vec{b}^{t,\theta}_i)_k)}{\sum^{N_s}_{i'=1}\exp((\vec{b}^{t,\theta}_{i'})_k)}.
\end{equation}
where $k\in\{1,2,\cdots,K\}$.

For each query $\hat{\vec{h}}^{t,\theta}_i$, we use the $\vec{v}^{t,\theta}$ to obtain aggregated memory features, which is computed as the weighted average over all memory items as $((\vec{v}^{t,\theta}_i)^T\vec{M}^\theta)^T$. Then, we output the globally refined feature $\vec{q}^{t,\theta}_i$ by combining the information in query $\hat{\vec{h}}^{t,\theta}_i$ (i.e. stock-concept feature) and aggregated items $((\vec{v}^{t,\theta}_i)^T\vec{M}^\theta)^T$ as an element-wise production, that is
\begin{equation}\label{eq323}
    \vec{q}^{t,\theta}_i=\hat{\vec{h}}^{t,\theta}_i\otimes((\vec{v}^{t,\theta}_i)^T\vec{M}^\theta)^T,
\end{equation}
where `$\otimes$' stands for element-wise production operator. Note that when refining the concept features in the global aggregation process, we used all the memory items, because we assume that, in this way, the model retrieves more comprehensive and diverse information from the memory.

\subsubsection{Memorization} 

Memory items need to be updated by the memory block during training. For each query $\hat{\vec{h}}^{t,\theta}_{i}$, the value to be updated is determined according to the matching probability in \eqref{eq321} obtained from the aggregation process. We renormalize the probabilities $\vec{v}^{t,\theta}_{i}$ for each stock $i$ in the updating process, denoted by $\hat{\vec{v}}^{t,\theta}_{i}$ as
\begin{equation} \label{eq326}
    (\hat{\vec{v}}^{t,\theta}_{i})_k=\frac{(\vec{v}^{t,\theta}_i)_k}{\max_{1\leq\ {k'} \leq K}\{(\vec{v}^{t,\theta}_{i})_{k'}\}}.
\end{equation}
For each stock $i$, we define a sum $s^{t,\theta}_i=\sum_{k=1}^{K}(\hat{\vec{v}}^{t,\theta}_{i})_k$. Then, we sort all stocks according their sum to obtain a decreasing sequence $\{i_1,i_2,\cdots,i_{N_s}\}$ such that $s^{t,\theta}_{i_1}\geq s^{t,\theta}_{i_2}\geq\cdots\geq s^{t,\theta}_{i_{N_s}}$. To update memory items, we select the top-$K$ stocks denoted by $\{i_1,i_2,\cdots,i_{K}\}$.

We denoted by $(\vec{M}^{\theta})_{k:} \in \mathbb{R}^{L}$ the $k$-th row of matrix $\vec{M}^{\theta}$. 
For each memory item $(\vec{M}^{\theta})_{k:}$, it will be updated by the $k$-th stock $i_k$ in above sequence $\{i_1,i_2,\cdots,i_{K}\}$. That is
\begin{equation}\label{eq327}
    (\vec{M}^{\theta})_{k:}\leftarrow L2\left((\vec{M}^{\theta})_{k:}+s^{t,\theta}_{i_k}(\hat{\vec{h}}^{t,\theta}_{i_k})^T\right)
\end{equation}
where $L2(\cdot)$ implies we use L2 normalization. Unlike the aggregating process, when absorbing current information into the memory items, we used only part of the queries that are most related to the items since we assume that this can filter out noise information in the dataset and make our model concentrate more on real profit signals.

\subsection{Stock Trend Regressor} \label{regressor}
For each module $\theta\in\{1,2,3\}$, after obtaining the corresponding stock-concept feature $\hat{\vec{h}}^{t,\theta}_i$ according to the above procedures, we can compute its forecast feature $\vec{y}^{t,\theta}_i$. Thus, we have
\begin{equation}\label{eq327}
    \vec{y}^{t,\theta}_i={\rm LeakyReLU}\left[\vec{W}_f\hat{\vec{h}}^{t,\theta}_i+\vec{b}_f\right],
\end{equation}
where $\vec{W}_f$ and $\vec{b}_f$ are learnable FC parameters.

Finally, all the forecast features will be collected by the stock trend regressor to make the forecast shown as the red block in Fig. \ref{fig:frame2}, that is
\begin{equation}
\hat{p}^t_i=\vec{W}_{p}\left(\vec{y}_i^{t,1}+\vec{y}_i^{t,2}+\vec{y}_i^{t,3}\right)+\vec{b}_p,
\end{equation}
where $\hat{p}^t_i$ is the predicted trend of stock $i$ on date $t$ and $W_p$, and $b_p$ are learnable parameters. The loss function of the whole model is formulated as
\begin{equation}
    \mathcal{L}=\sum\nolimits_{t \in T}{\rm MSE}\left(\hat{\vec{p}}^t, \vec{p}^t\right)=\sum\nolimits_{t \in T}\sum\nolimits^{N_s}_{i=1}\frac{(\hat{p}^t_i-p^t_i)^2}{N_s},
\end{equation}
where $T$ is the set of dates included in the training set, $p^t_i$ is the ground truth of the trend of stock $i$ on date $t$, and ${\rm MSE}$ denotes the mean square error.

\section{Experiments}\label{Exp}
In this section, we first compare the performance of our framework with other key related work and then test the effect of our framework in different blocks through an ablation study. The results show that the memory module fits MTMD best when it is used to refine stock-concept features obtained by local aggregation, and this is effective for both predefined and hidden concept blocks. Experimental results and details are shown in the subsections.

\subsection{Datasets and Experimental Settings}

\subsubsection{Datasets} The source of the dataset is CSI 100 and CSI 300. The CSI 300 comprises the 300 most representative securities with large scale and good liquidity in the Shanghai and Shenzhen markets, which can reflect the overall performance of markets. The top three industries are industrial, financial, and major consumer. CSI 100 selects the most extensive 100 stocks in the CSI 300 to comprehensively reflect the overall situation of the large-cap companies with the most vital market influence. Therefore, they can represent the situation of the entire A-share market.

The stock features come from Alpha360 of the quantitative investment platform Qlib \cite{yang2020qlib}. This dataset reviews the basic trading information of stocks in the past 60 days as a feature of the stock on that date, which includes opening price, closing price, high price, low price, volume-weighted average price (VWAP), trading volume, target stocks including industry and the main business. The average number of target stocks on each day is 410 in CSI 100 and 1344 in CSI 300. The time span included in this dataset is from 01/01/2007 to 12/31/2020, with training set from 01/01/2007 to 12/31/2014, validation set from 01/01/2015 to 12/31/2016, and test set from 01/01/2017 to 12/31/2020. The forecast target is the daily change rate of each stock defined as $p_i^t=\frac{price^{t+1}-price^{t}}{price^{t}}$ and the change rates on the same date are normalized during data preprocessing.

\subsubsection{Baselines} 
We compare our MTMD framework with the following models:

\begin{itemize}
\item \textbf {GATs \cite{casanova2018graph}:}  The graph attention network (GAT) is a network architecture with shielded self-attention layers used to solve the shortcomings of traditional graph convolution or other similar existing approaches.

\item \textbf {MLP:} The multi-layer perceptron (MLP) is comprised of 3 linear layers with 512 units for each layer.

\item \textbf {SFM \cite{zhang2017stock}:} The novel state frequency memory (SFM) cyclic network captures multi-frequency trading patterns to make long-term and short-term forecasts from past market data.

\item \textbf {GRU \cite{chung2014empirical}:} The gated recurrent unit (GRU) network consists of two cycles. The encoder and decoder of the proposed model are jointly trained to maximize the conditional probability series of the target sequence of a given source.

\item \textbf {LSTM \cite{hochreiter1997long}:} The long short-term memory (LSTM) network used minimum time lag of bridging discrete time steps, which was learned by forcing a constant error stream through constant error rotation in a special model.

\item \textbf {ALSTM \cite{feng2018enhancing}:} The recurrent neural network is based on two-stage attention (DA-RNN), which aims to solve the problem of time dependence and correlation.

\item \textbf {ALSTM+TRA \cite{lin2021learning}:} In addition to modeling multiple trading patterns, The ALSTM is extended with a Temporal Routing Adaptor (TRA).

\item \textbf {HIST \cite{xu2021hist}:} A graph-based framework for stock trend forecasting via mining concept-oriented Shared Information (HIST) can fully extract shared information from graphs constructed with predefined concepts and hidden concepts to improve the performance of stock trend forecast.

\item \textbf{PatchTST \cite{PatchTST}}: The Patch Time Series Transformer (PatchTST) introduces an efficient design for multivariate time series forecasting, segmenting time series into subseries-level patches as input tokens and adopting a channel-independent approach. This model retains local semantic information, reduces computation and memory usage, and improves forecasting accuracy, especially for long-term predictions.

\item \textbf{iTransformer \cite{liu2023itransformer}:} The iTransformer repurposes the Transformer architecture without modifying its components, applying attention and feed-forward networks on inverted dimensions to learn variate-centric representations. This approach enhances performance by capturing multivariate correlations and utilizing arbitrary lookback windows effectively, achieving state-of-the-art results on real-world datasets.

\item \textbf{MASTER \cite{li2024master}:} The Market-Guided Stock Transformer (MASTER) models momentary and cross-time stock correlations while leveraging market information for automatic feature selection. It alternates between intra-stock and inter-stock aggregation to handle complex stock correlations, providing superior forecasting performance and insightful visualization of realistic stock interactions.
\end{itemize}

\begin{table}[!t]
    \caption{The selection of the hyper-parameters.}
    \resizebox{\linewidth}{!}{
	\begin{tabular}{l|ccc|ccc}
		\hline
		Model &\multicolumn{3}{|c|}{Numbers of Units} &\multicolumn{3}{|c}{Numbers of Layers} \\
		 & CSI 100 & CSI 300 & CSI 500 & CSI 100 & CSI 300 & CSI 500  \\ \hline
		GATs & 128 & 64 & 64 & 2 & 2& 3 \\
		MLP & 512 & 512& 128 & 3 & 3& 3 \\
		SFM & 64 & 128& 64 & 2 & 2& 2 \\
		GRU & 128 & 64& 64 & 2 & 2& 2 \\
		LSTM & 128 & 64& 128 & 2 & 2& 3 \\
		ALSTM & 64 & 128& 64 & 2 & 2& 2 \\
		ALSTM+TRA & 64 & 128& 64 & 2 & 2& 3 \\
		HIST & 128 & 128& 64 & 2 & 2& 1 \\
		PatchTST & 128 & 64& 128 & 2 & 2& 1 \\
		iTransformer & 64 & 256& 128 & 2 & 2& 1 \\
		MASTER & 256 & 128& 64 & 3 & 2& 2 \\ \hline
		{\textbf{MTMD}} & {\textbf{128}} & {\textbf{128}}& {\textbf{128}} & {\textbf{2}} & {\textbf{2}}& {\textbf{2}} \\ 
		\hline
	\end{tabular}}
	\centering
\label{HH}
\end{table}

\subsection{Contrast Experiments}

\begin{table*}[!t]
\centering
\caption{The main results (and their standard deviations) on CSI 100 and CSI 300.}
    \resizebox{\linewidth}{!}{
	\begin{tabular}{c|cc|cccc|cc|cccc}
		\hline
		\multirow{3}{*}{Methods}
		& \multicolumn{6}{c|}{CSI 100} & \multicolumn{6}{c}{CSI 300} \\
		\cline{2-13}
		
		& IC & Rank IC & \multicolumn{4}{c|}{Precision@N ($\uparrow$)} & IC & Rank IC & \multicolumn{4}{c}{Precision@N ($\uparrow$)} \\ 
		 
		 & ($\uparrow$)&($\uparrow$) &3 &5 &10 &30 & ($\uparrow$) & ($\uparrow$) &3 &5 &10 &30 \\ 
		 \hline
		 
		\multirow{2}{*}{MLP}
		& 0.071 & 0.067 & 56.53 & 56.17 & 55.49 & 53.55 & 0.082 & 0.079 & 57.21 & 57.10 & 56.75 & 55.56  \\
		& (4.8e-3) & (5.2e-3) & (0.91) & (0.48) & (0.30) & (0.36) & (6.0e-4) & (3.0e-4) & (0.39) & (0.33) & (0.34) & (0.14) \\\hline

		\multirow{2}{*}{SFM \cite{zhang2017stock}}                                             
		& 0.081 &0.074 &57.79 &56.96 &55.92 &53.88 &0.102 &0.096 &59.84 &58.28 &57.89 &56.82 \\ 
		& (7.0e-3)  & (8.0e-3) & (0.76)  & (1.04) & (0.60) & (0.47) & (3.0e-3) & (2.7e-3) & (0.91) & (0.42) & (0.45) & (0.39) \\\hline
		
		\multirow{2}{*}{GATs \cite{casanova2018graph}}             
		& 0.096 &0.090 &59.17 &58.71 &57.48 &54.59 &0.111 &0.105 &60.49 &59.96 &59.02 &57.41 \\
		& (4.5e-3) & (4.4e-3) & (0.68) & (0.52) & (0.30) & (0.34)  & (1.9e-3) & (1.9e-3) & (0.39) & (0.23) & (0.14) & (0.30) \\\hline
		
		\multirow{2}{*}{LSTM \cite{hochreiter1997long}}               
		& 0.097 &0.091 &60.12 &59.59 &59.04 &54.77 &0.104 &0.098 &59.51 &59.27 &58.40 &56.98 \\
		& (2.2e-3)  & (2.0e-3) & (0.52) & (0.19) & (0.15) & (0.11) & (1.5e-3) & (1.6e-3) & (0.46) & (0.34) & (0.30) & (0.11) \\ \hline
		
		\multirow{2}{*}{ALSTM \cite{feng2018enhancing}}                                                 
		& 0.102 &0.094 &60.79 &59.76 &58.13 &55.00 &0.115 &0.109 &59.51 &59.33 &58.92 &57.47 \\
		& (1.8e-3) & (1.9e-3) & (0.23) & (0.42) & (0.13) & (0.12) & (1.4e-3) & (1.4e-3) & (0.20) & (0.51) & (0.29) & (0.16) \\ \hline
		
		\multirow{2}{*}{GRU \cite{chung2014empirical}}                                                 
		& 0.103 & 0.097 &59.97 &58.99 &58.37 & 55.09 &0.113 &0.108 &59.95 &59.28 &58.59 &57.43 \\ 
		& (1.7e-3) & (1.6e-3) & (0.63)  & (0.42) & (0.29) & (0.15) & (1.0e-3) & (8.0e-4) & (0.62) & (0.35) & (0.40) & (0.28) \\\hline
		
		\multirow{2}{*}{ALSTM+TRA \cite{lin2021learning}}                                                 
		& 0.107 &0.102 &60.27 &59.09 &57.66 &55.16 &0.119 &0.112 &60.45 &59.52 &59.61 &58.24 \\
		& (2.0e-3)  & (1.8e-3) & (0.43) & (0.42) & (0.33) & (0.22) & (1.9e-3) & (1.7e-3) & (0.53) & (0.58) & (0.43) & (0.32)\\ \hline
		
		\multirow{2}{*}{HIST \cite{xu2021hist}}                                                 
		& 0.120 &0.115 &61.87 &60.82 &59.38 &56.04 &0.131 &0.126 &61.60 &61.08 &60.51 &58.79 \\
		& (1.7e-3) & (1.6e-3) & (0.47) & (0.43) & (0.24) & (0.19) & (2.2e-3) & (2.2e-3) & (0.59) & (0.56) & (0.40) & (0.31) \\ \hline

		\multirow{2}{*}{PatchTST \cite{PatchTST}}
		& 0.118 & 0.113 &61.50 &60.50 &59.00 &56.00 & 0.129 & 0.124 &61.30 &60.80 &60.00 &58.50 \\
		& (1.5e-3) & (1.4e-3) & (0.40) & (0.35) & (0.20) & (0.15) & (1.3e-3) & (1.2e-3) & (0.50) & (0.40) & (0.30) & (0.25) \\ \hline

	\multirow{2}{*}{iTransformer \cite{liu2023itransformer}}
		& 0.119 & 0.114 &61.60 &60.70 &59.20 &56.20 & 0.130 & 0.125 &61.50 &61.00 &60.20 &58.70 \\
		& (1.4e-3) & (1.3e-3) & (0.38) & (0.33) & (0.18) & (0.14) & (1.2e-3) & (1.1e-3) & (0.48) & (0.38) & (0.28) & (0.23) \\ \hline

	\multirow{2}{*}{MASTER \cite{li2024master}}
		& 0.122 & 0.117 &61.31 &60.11 &59.10 &55.30 & 0.133 & 0.128 &62.00 &61.50 &60.80 &59.00 \\
		& (1.3e-3) & (1.2e-3) & (0.37) & (0.32) & (0.17) & (0.13) & (1.1e-3) & (1.0e-3) & (0.47) & (0.37) & (0.27) & (0.22) \\ \hline

		\multirow{2}{*}{\textbf{MTMD}}  
		&\textbf{0.128} & \textbf{0.120} & \textbf{61.62} & \textbf{60.89} & \textbf{59.70} & \textbf{56.25} & \textbf{0.138} & \textbf{0.133} & \textbf{62.51} & \textbf{61.61} & \textbf{60.80} & \textbf{59.25} \\
		& \textbf{(2.0e-3)} & \textbf{(1.3e-3)} & \textbf{(0.66)} & \textbf{(0.53)} & \textbf{(0.08)} & \textbf{(0.19)} & \textbf{(1.4e-3)} & \textbf{(1.3e-3)} & \textbf{(0.55)} & \textbf{(0.35)} & \textbf{(0.17)} & \textbf{(0.11)} \\ \hline
	\end{tabular}}
\label{Experiment Results1}
\end{table*}

\begin{table}[!t]
\centering
\caption{The main results (and their standard deviations) on CSI 500.}
    \resizebox{\linewidth}{!}{
	\begin{tabular}{c|cc|cccc}
		\hline
		\multirow{3}{*}{Methods}
		& \multicolumn{6}{c}{CSI 500} \\
		\cline{2-7}
		
		& IC & Rank IC & \multicolumn{4}{c}{Precision@N ($\uparrow$)} \\ 
		 
	    & ($\uparrow$)&($\uparrow$)&3 &5 &10 &30\\ 
		\hline
	
		\multirow{2}{*}{MLP}
	    & 0.085 & 0.081 & 57.50 & 57.30 & 56.90 & 55.80 \\
	  & (5.0e-4) & (4.0e-4) & (0.40) & (0.35) & (0.30) & (0.15) \\\hline

		\multirow{2}{*}{SFM}                                             
		& 0.105 & 0.098 & 60.10 & 58.50 & 58.00 & 57.00 \\ 
		& (2.5e-3) & (2.5e-3) & (0.80) & (0.50) & (0.40) & (0.35) \\\hline
		
		\multirow{2}{*}{GATs}             
		& 0.114 & 0.108 & 60.80 & 60.20 & 59.30 & 57.80 \\
		& (1.8e-3) & (1.8e-3) & (0.60) & (0.40) & (0.20) & (0.25) \\\hline
		
		\multirow{2}{*}{LSTM}               
		& 0.106 & 0.100 & 59.80 & 59.50 & 58.70 & 57.20 \\
		& (1.4e-3) & (1.5e-3) & (0.50) & (0.30) & (0.25) & (0.10) \\ \hline
		
		\multirow{2}{*}{ALSTM}                                                 
		& 0.117 & 0.110 & 60.00 & 59.60 & 59.00 & 58.00 \\
		& (1.3e-3) & (1.3e-3) & (0.30) & (0.40) & (0.20) & (0.15) \\ \hline
		
		\multirow{2}{*}{GRU}                                                 
		& 0.115 & 0.110 & 60.10 & 59.50 & 58.80 & 57.70 \\ 
		& (1.2e-3) & (1.1e-3) & (0.60) & (0.40) & (0.35) & (0.20) \\\hline
		
		\multirow{2}{*}{ALSTM+TRA}                                                 
		& 0.121 & 0.115 & 60.70 & 60.00 & 59.80 & 58.50 \\
		& (1.5e-3) & (1.6e-3) & (0.50) & (0.45) & (0.30) & (0.25) \\ \hline
		
		\multirow{2}{*}{HIST}                                                 
		& 0.133 & 0.128 & 61.90 & 61.20 & 60.70 & 59.00 \\
		& (1.8e-3) & (1.8e-3) & (0.50) & (0.40) & (0.35) & (0.28) \\ \hline

		\multirow{2}{*}{PatchTST}
		& 0.131 & 0.126 & 61.50 & 60.90 & 60.20 & 58.70 \\
		& (1.2e-3) & (1.1e-3) & (0.45) & (0.35) & (0.25) & (0.20) \\ \hline

		\multirow{2}{*}{iTransformer}
		& 0.132 & 0.127 &61.70 &61.10 &60.40 &58.90 \\
		& (1.1e-3) & (1.0e-3) & (0.43) & (0.33) & (0.23) & (0.18) \\ \hline

		\multirow{2}{*}{MASTER}
		& 0.135 & 0.130 &62.10 &61.50 &60.90 &59.20 \\
		& (1.0e-3) & (9.0e-4) & (0.42) & (0.32) & (0.22) & (0.17) \\ \hline

		\multirow{2}{*}{\textbf{MTMD}}  
		& \textbf{0.140} & \textbf{0.135} & \textbf{62.80} & \textbf{62.00} & \textbf{61.00} & \textbf{59.50} \\
		& \textbf{(1.3e-3)} & \textbf{(1.2e-3)} & \textbf{(0.50)} & \textbf{(0.40)} & \textbf{(0.20)} & \textbf{(0.15)} \\ \hline
	\end{tabular}}
\label{Experiment Results2}
\end{table}

\subsubsection{Experimental Settings}
The evaluation metrics used in our experiment are the Information Coefficient (IC), Rank Information Coefficient (Rank IC), and Precision@N. 

\begin{itemize}
\item \textbf {IC:} The IC is defined as the correlation between predicted and actual values, serving as a metric to assess the predictive accuracy and stock selection capability of a model. Formally, the IC, constrained within the interval $[-1, 1]$, quantifies the linear correlation between the rankings of all stocks at the beginning of an adjustment cycle and their return rankings at the cycle's conclusion. The mathematical representation of IC is given by
\begin{equation}
\text{IC} = \frac{1}{N} \frac{(\hat{y}-\text{mean}(\hat{y}))^{\top}(y-\text{mean}(y))}{\text{std}(\hat{y}) \times \text{std}(y)},
\end{equation}
where $\hat{y}$ denotes the predicted stock position weight rankings, and $y$ represents the actual stock return rankings. In practical applications, an IC exceeding 0.05 signifies an effective factor, whereas an IC surpassing 0.1 denotes a robust alpha factor. Conversely, an average IC value approximating zero indicates the factor's ineffectiveness.

\item \textbf {Rank IC:} The Rank IC is calculated as the correlation coefficient between the rank of factor loadings in period $t$ and the rank of factor returns in period $t+1$. The expression for Rank IC is
\begin{equation}
\text{Rank IC} = \text{corr}(\mathbf{r}_{factor}^{t}, \mathbf{r}_{return}^{t+1}),
\end{equation}
where $\text{corr}(\cdot)$ represents the rank correlation coefficient, $\mathbf{r}_{factor}$ denotes the factor loading rankings, and $\mathbf{r}_{return}$ indicates the return rankings of the factor.

\item \textbf {Precision@N:} The Precision@N is the proportion of top N forecasts on each day with positive labels. In order to evaluate the precision comprehensively, we use different N values, including 3, 5, 10, and 30. 
\end{itemize}

All results of metrics are presented as the average value of these metrics on each day. The programming implementation of our algorithm and framework is mainly based on the PyTorch library \cite{paszke2019pytorch}, and all experiments were run on a single NVIDIA Tesla A100 GPU.

\subsubsection{Hyper-Parameters} 
Since the training and test batches for Alpha360's stock features come from the same date and have 360-degree dimensions, the batch size is decided by the number of stocks traded each day. In addition, the shape of MTMD's memory matrix is $64\times 64$. We also use gird research to find optimal hyper-parameters. Table \ref{HH} displays the selection of the number of units and layers in other baselines and MTMD. By looking up the hyper-parameters in the stock features encoder for MTMD, the number of hidden units on CSI 500, CSI 100 and CSI 300 are both 128 in $\left\{32,64,128,256,512 \right\}$, the number of layers and the shape of MTMD's memory matrix are 2 in $\left\{1,2,3,4 \right\}$ and $64\times 64$ in $\left\{32, 64,128,256 \right\}$. The best learning rate is 0.0002 in $\left \{0.001, 0.0005, 0.0002, 0.0001 \right \}$.

Given the same dataset and predefined concepts, Table \ref{Experiment Results1} shows the experimental results of MTMD and other baselines on the stocks of CSI 100 and CSI 300. Our framework achieves state-of-the-art IC, Rank IC, and Precision@N, which can be further validated in dataset CSI 500 as shown in Table \ref{Experiment Results2}. The CSI 500 is a supplementary dataset that includes 500 constituent stocks representing mid-cap companies on the Chinese stock market. This dataset extends the evaluation to a broader range of stocks, covering more diverse market conditions and trading patterns, thereby providing a more comprehensive validation of the model’s generalization capabilities.

\begin{table*}[!t]
\caption{The results of ablation study on CSI 100.}
\centering
    \resizebox{\linewidth}{!}{
    \begin{tabular}{c|cc|cccc|c}
        \hline
        Methods & IC ($\uparrow$) & Rank IC ($\uparrow$) & Precision@3 ($\uparrow$) & Precision@5 ($\uparrow$) & Precision@10 ($\uparrow$) & Precision@30 ($\uparrow$) & Memory Module Enabled \\ \hline

        \multirow{1}{*}{HIST}
        & 0.120 & 0.120 & 61.62 & 60.89 & 58.70 & 56.25 & Baseline (B)\\ \hline
        \multirow{2}{*}{MTMD}
        & 0.124 & 0.118 & 62.86 & 61.62 & 59.68 & 56.27 & \multirow{2}{*}{Predefined Concept Block (P)}\\
        & (4.7e-4) & (9.4e-4) & (0.74) & (0.42) & (0.33) & (0.21) &  \\ \hline
        \multirow{2}{*}{MTMD}
        & 0.126 & 0.120 & 62.24 & 61.13 & 59.86 & 56.17 & \multirow{2}{*}{Hidden Concept Block (H)}\\
        & (2.0e-3) & (2.9e-3) & (0.13) & (0.30) & (0.49) & (0.15) &  \\ \hline
        \multirow{2}{*}{MTMD}
        & 0.128 & 0.120 & 61.62 & 60.89 & 59.70 & 56.25 & \multirow{2}{*}{All (A)}\\ 
        & (2.0e-3) & (1.3e-3) & (0.66) & (0.53) & (0.08) & (0.19) &  \\ \hline
    \end{tabular}}
\label{Experiment Results2}
\end{table*}

\begin{table*}[!t]
\caption{The results of ablation study on CSI 300.}
\centering
    \resizebox{\linewidth}{!}{
    \begin{tabular}{c|cc|cccc|c}
        \hline
        Methods & IC ($\uparrow$) & Rank IC ($\uparrow$) & Precision@3 ($\uparrow$) & Precision@5 ($\uparrow$) & Precision@10 ($\uparrow$) & Precision@30 ($\uparrow$) & Memory Module Enabled \\ \hline

        \multirow{2}{*}{HIST}
        & 0.131 & 0.126 & 61.60 & 61.08 & 60.51 & 58.79 & 
        \multirow{2}{*}{Baseline (B)}\\
        & (2.2e-3) & (2.2e-3) & (0.59) & (0.56) & (0.40) & (0.31) &  \\ \hline
        
        \multirow{2}{*}{MTMD}
        & 0.136 & 0.132 & 61.31 & 61.19 & 61.12 & 59.46 & \multirow{2}{*}{Predefined Concept Block (P)}\\
        & (3.0e-3) & (3.6e-3) & (0.26) & (0.07) & (0.56) & (0.46) &  \\ \hline
        \multirow{2}{*}{MTMD}
        & 0.137 & 0.132 & 62.68 & 61.65 & 60.76 & 59.15 & \multirow{2}{*}{Hidden Concept Block (H)}\\
        & (4.7e-4) & (4.7e-4) & (0.64) & (0.28) & (0.22) & (0.27) &  \\ \hline
        \multirow{2}{*}{MTMD}
        & 0.138 & 0.133 & 62.51 & 61.61 & 60.80 & 59.25 & \multirow{2}{*}{All (A)}\\
        & (1.4e-3) & (1.3e-3) & (0.55) & (0.35) & (0.17) & (0.11) &  \\ \hline
    \end{tabular}}
\label{Experiment Results3}
\end{table*}

Compared with the SFM, ALSTM, ALSTM+TRA, and GRU, which predict stock trends through time series, our framework pays more attention to mining the conceptual features of stocks. The IC of MTMD is 0.008 higher than the best score of ALSTM+TRU on average. This result shows that the concept of stock can better mine the correlation of cross-stock changes to achieve a better forecast performance. Besides, although GATs can also capture cross-stock connections, it does not study the dynamic relevance degree between stocks and concepts. However, HIST can more effectively capture the temporal and complicated cross-stock relations between stocks. The performance is higher than the IC of GATs, with an average value of 0.112. This means that dynamically correlating temporal features with conceptual features can more accurately predict stock trends.


\begin{figure}[!t]
    \centering
    \includegraphics[width=0.8\linewidth]{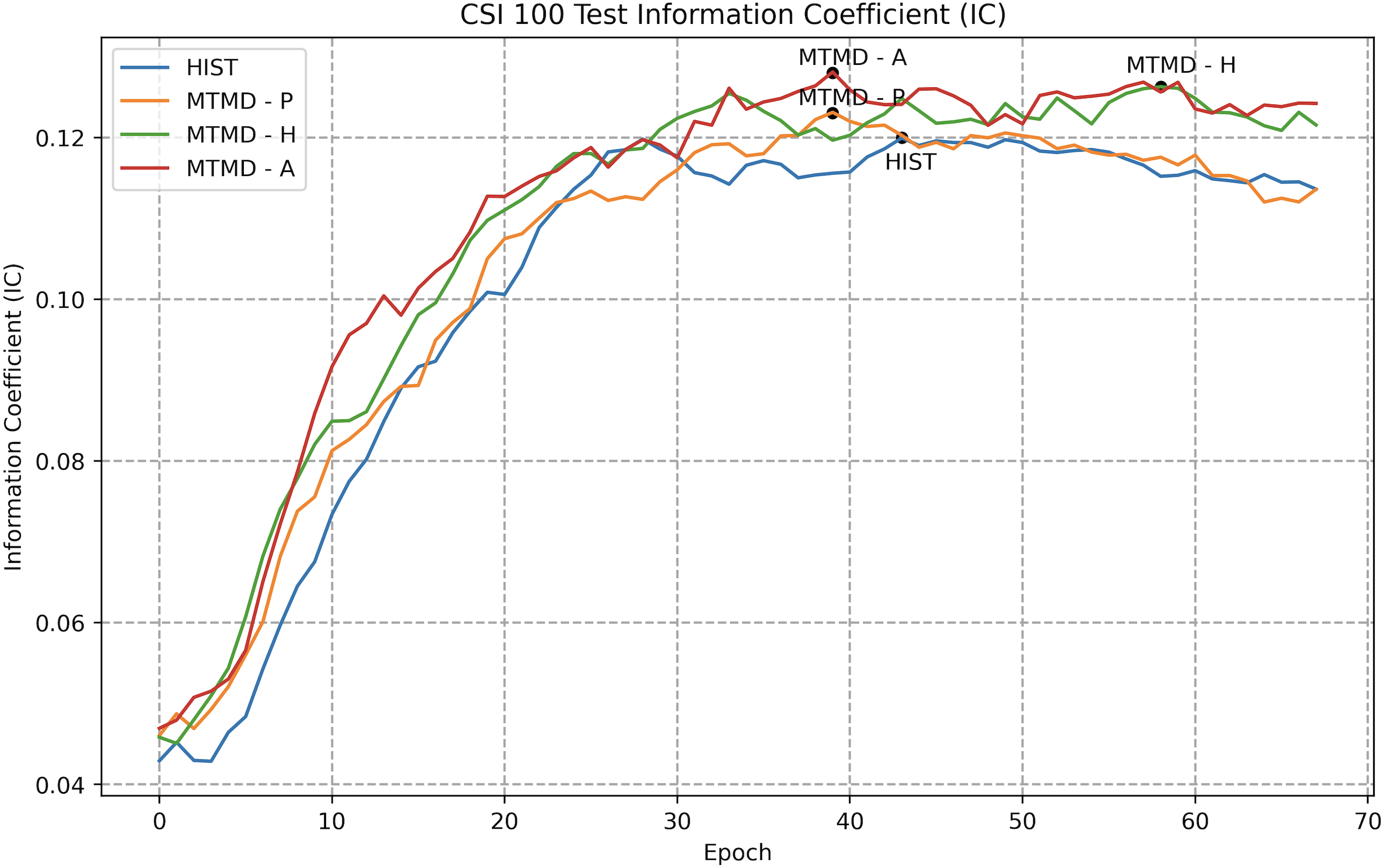}
    \caption{Ablation study of IC on CSI 100.}
    \label{FIGURE5}
\end{figure}

\subsection{Ablation Study}
In the ablation study, we test the significance of the memory module. We adopt HIST as our baseline, which only locally aggregates stock and concept information in each time step. In detail, our framework uses a 2-layer GRU network to encode the time-series features of several target stocks, plus a multi-scale aggregating process with the memory module for predefined and hidden concept blocks, respectively. The experiment results are shown in Table \ref{Experiment Results2} and Table \ref{Experiment Results3}. We will then demonstrate the effect of the global aggregation with memory items for each block.

\begin{table*}[!t]
\caption{The results of universality evaluation on CSI 300.}
\centering
    \resizebox{\linewidth}{!}{
    \begin{tabular}{c|cc|cccc}
        \hline
        Methods & IC ($\uparrow$) & Rank IC ($\uparrow$) & Precision@3 ($\uparrow$) & Precision@5 ($\uparrow$) & Precision@10 ($\uparrow$) & Precision@30 ($\uparrow$) \\ \hline

        \multirow{2}{*}{HIST}
        & 0.131 & 0.126 & 61.60 & 61.08 & 60.51 & 58.79 \\
        & (2.2e-3) & (2.2e-3) & (0.59) & (0.56) & (0.40) & (0.31) \\   \hline

        \multirow{2}{*}{MTMD-HIST}
        & 0.138 & 0.133 & 62.51 & 61.61 & 60.80 & 59.25 \\
        & (1.4e-3) & (1.3e-3) & (0.55) & (0.35) & (0.17) & (0.11) \\   \hline
        
        \multirow{2}{*}{MLP}
        & 0.082 & 0.079 & 57.21 & 57.10 & 56.75 & 55.56 \\
        & (6.0e-4) & (3.0e-4) & (0.39) & (0.33) & (0.34) & (0.14) \\   \hline
        
        \multirow{2}{*}{MTMD-MLP}
        & 0.085 & 0.082 & 57.85 & 57.89 & 57.30 & 56.50 \\
        & (1.2e-3) & (1.2e-3) & (1.21) & (0.17) & (0.28) & (0.23) \\   \hline

    \end{tabular}}
\label{ob}
\end{table*}

We first add a memory module to the predefined and hidden concept block separately. Then, when the memory module is applied to the predefined concept block, for test metrics of our model over CSI 100 and CSI 300 datasets, IC is improved by 0.006, Rank IC is improved by 0.006, and Precision@N rises by 0.694, on average. Also, the application of the memory module in the hidden concept block shows that IC is improved by 0.007, and Precision@N rises by 0.763, for the average value of metrics over CSI 100 and CSI 300 datasets. The better results show that information recorded in memory items is conducive to refining locally aggregated stock-concept features.

Moreover, we apply the memory module on these two blocks simultaneously with two different sets of memory items for each block. For average metrics over CSI 100 and CSI 300 datasets, IC is improved by 0.009, Rank IC rises by 0.007, and Precision@N increases by 0.437. From Fig. \ref{FIGURE5}, it can be observed that the performance is improved more obviously, which means that the simultaneous application of the memory module on these two concept blocks will not cause conflicts, and the overall performance will get promoted.

Therefore, the application of multi-scale aggregation can help the model to capture actual profit signals through different time and inventory levels, and this increases the model's robustness against external noise.

\begin{figure}[!t]
	\centering
	\subfigure[Global Aggregation in Predefined Concept Block]{
		\includegraphics[width=0.8\linewidth]{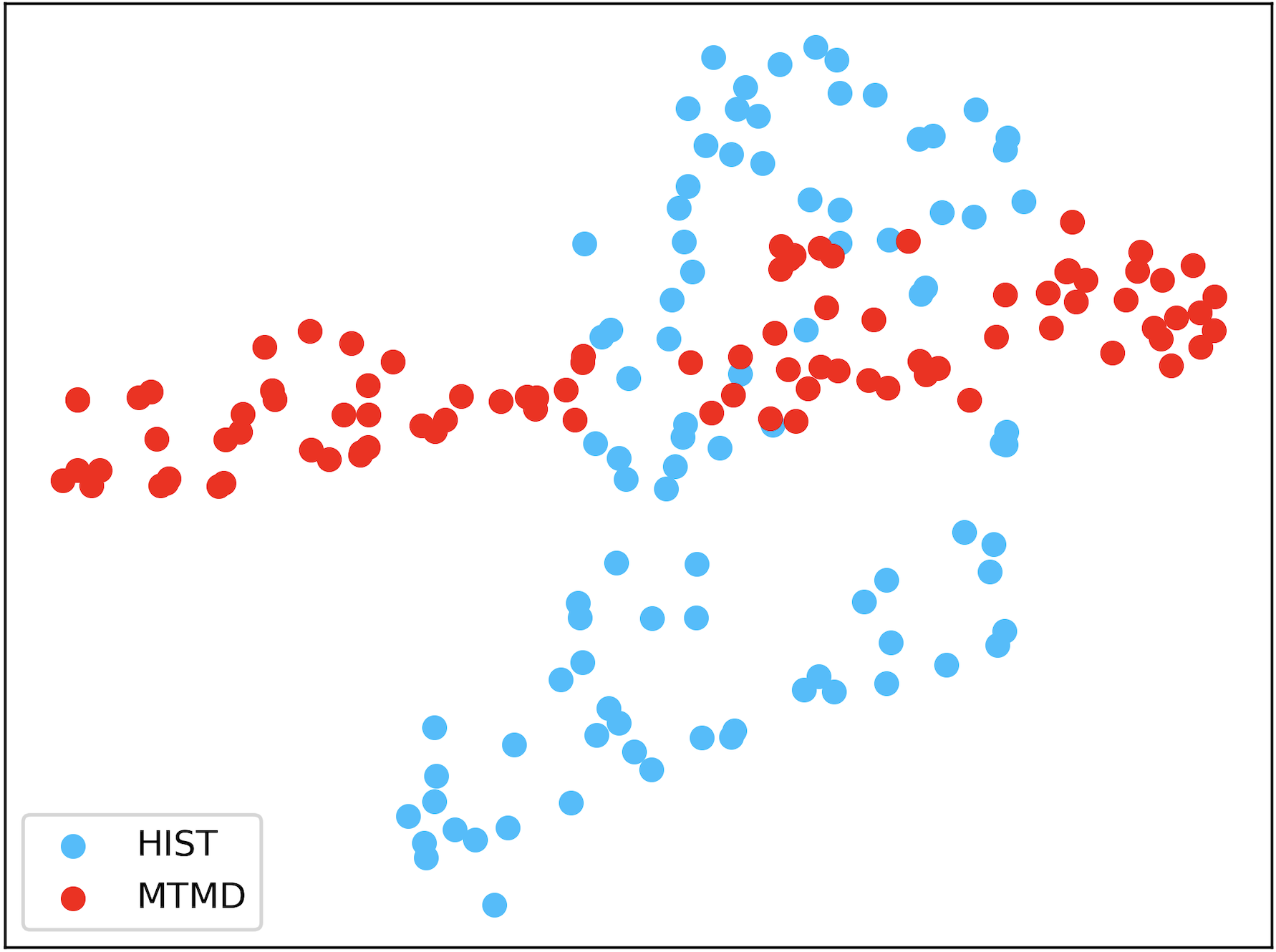}
            \label{Fig.sub.1}
	}%
	
	\subfigure[Global Aggregation in Hidden Concept Block]{
		\includegraphics[width=0.8\linewidth]{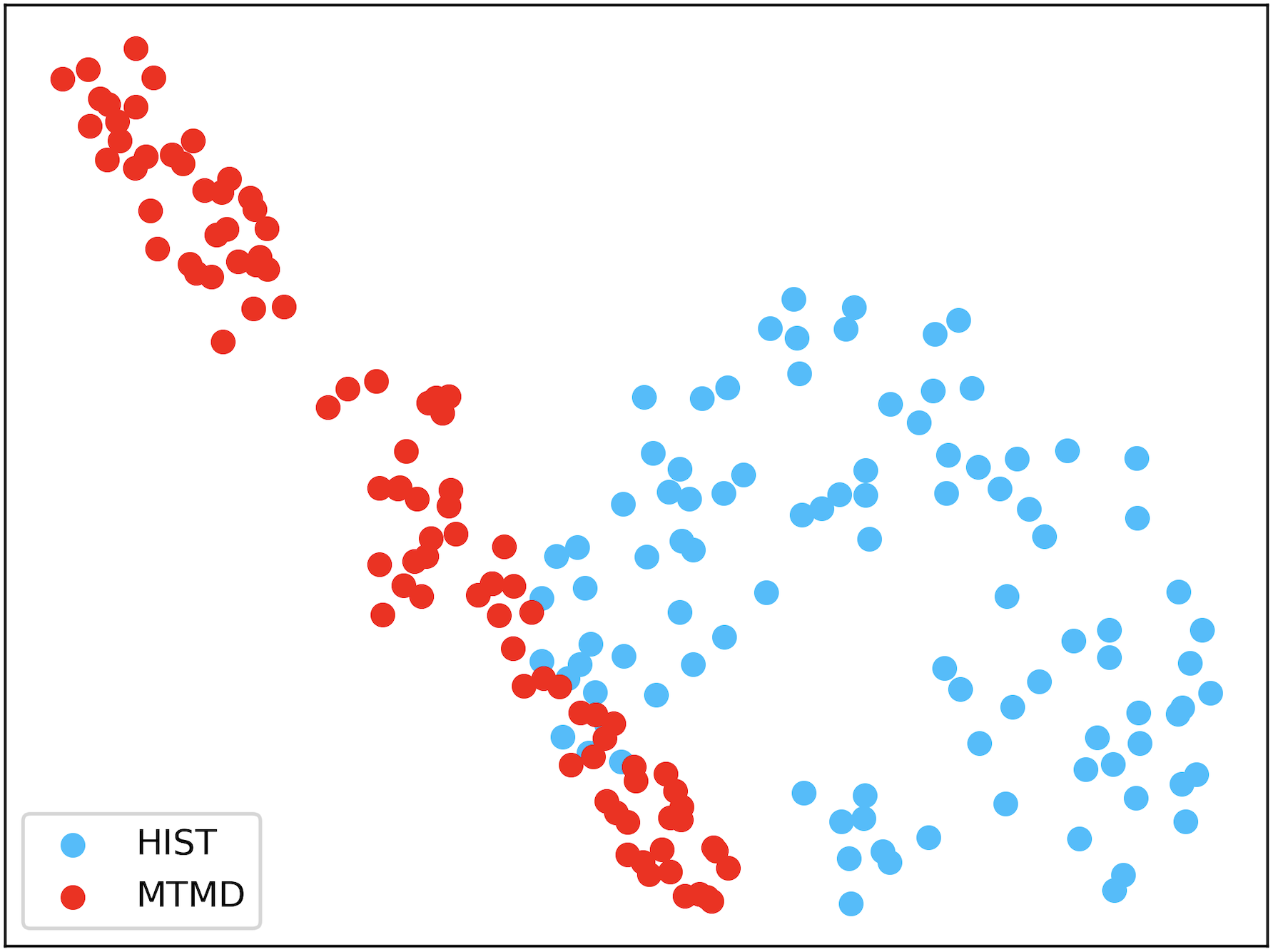}
            \label{Fig.sub.2}
	}%
	\caption{Visualization of memory matrix with base on t-SNE. And \textit{MTMD} can capture the real profit signal from the noise more effectively.}
	\label{Visualization of Debias}
\end{figure}

\subsection{Visualization of Debias}
We visualize the distribution of MTMD and HIST features based on t-SNE \cite{van2008visualizing} in Fig. \ref{Visualization of Debias}, with samples randomly chosen from CSI 300 dataset. We compare the distribution of stock-concept features of HIST and MTMD to interpret the effect of debiasing. The distribution of features generated by HIST is sparsely distributed in the dimension space, which means that the feature is noisy, and it is hard for the model to discriminate significant profit signals. With global aggregation performed by the memory module, the features obtained from MTMD are more densely distributed and show an apparent linear pattern. Moreover, the features generated by MTMD present clusters with clear margins indicating the disparity between different kinds of samples, while the features output by HIST scatter in the dimension space without an obvious clustering pattern. Hence, the memory module's debiasing effect is proved, making MTMD capture more meaningful information through the noisy stock market.

\subsection{Universality Evaluation} 
To validate the universality and plug-and-play nature of our method, we substituted the HIST backbone with an MLP. A memory-aware feature aggregator was incorporated before the classification head of the MLP model. The experimental outcomes are presented accordingly. The outcomes of these experiments, detailed in Table \ref{ob}, offer compelling evidence of the enhanced performance and versatility afforded by our method. This not only underscores the effectiveness of incorporating memory-aware mechanisms into traditional neural networks for stock trend forecasting but also illustrates the method's adaptability to different architectural foundations.

\begin{table}[!t]
    \caption{The results of efficiency evaluation for different models and components.}
    \centering
    {
    \begin{tabular}{c|c}
        \hline
        \textbf{Model/Component} & \textbf{Average Time (µs per sample)} \\ \hline
        HIST Model & Inference: 2780.97 \\ \hline
        3-Layer MLP & Inference: 162.73 \\ \hline
        Local Aggregation & Inference: 73.70 \\ \hline
        Global Aggregation & Inference: 74.99 \\ \hline

    \end{tabular}}
\label{tab:efficiency_evaluation}
\end{table}

\subsection{Efficiency Evaluation} 
To evaluate the computational efficiency of our proposed MTMD framework, we conducted a comprehensive analysis of the inference time of each component, particularly focusing on the local and global aggregation modules. The efficiency evaluation was conducted using the same hardware and software environment as the main experiments. We measured the average computation time per sample for key operations, including the read and upload functions within the local and global aggregation processes. For comparison, we also evaluated the inference time of two baseline models: the HIST model and a simple 3-layer MLP, which represent a graph-based approach and a lightweight neural network, respectively. 

As shown in Table \ref{tab:efficiency_evaluation}, The results indicate that the read operations within both local and global aggregation modules are extremely efficient, with average inference times below 75 $\mu s$ per sample. These operations are primarily used during the inference phase, ensuring that the overall latency introduced by the aggregation processes remains minimal.

\section{Conclusion}\label{conclusion}
In this paper, we propose a framework named Multi-scale Temporal Memory Learning and Efficient Debiasing (MTMD). With the help of memory items, our model not only aggregates stock-concept information effectively from a local scale but also judges the stock trend based on memorized real profit signals in recorded historical patterns. Specifically, we further enhance the temporal consistency via self-similarity, free from time intervals. Moreover, the complexity would decrease from $\mathcal{O}(T^2)$ (Recurrent Module) to $\mathcal{O}(T)$ (MTMD). The experimental results from stock sets CSI 100, CSI 300 and CSI 500 show that our model has reached new state-of-the-art performance and significantly improved the accuracy of stock trends. Despite our superior performance under real stock sets, markets are constantly changing. In future works, we will deploy our model under a more up-to-date stock market and make it more adaptive to more complex trading environments in real scenarios.


\section*{Acknowledgement}
This work was supported in part by the National Natural Science Foundation of China (NSFC) under Grant No. 62202055 and No. 61872239, the Interdisciplinary Incubation Project of Beijing Normal University under Grant 310430014; the Start-up Fund from Beijing Normal University under Grant No. 310432104/312200502510, the Internal Fund from BNU-HKBU United International College under Grant No. UICR0400003-24, the Project of Young Innovative Talents of Guangdong Education Department under Grant No. 2022KQNCX102, the Zhuhai Science-Tech Innovation Bureau under Grants No. ZH22017001210119PWC and No. 28712217900001, and the Interdisciplinary Intelligence Supercomputer Center, Beijing Normal University (Zhuhai).


\ifCLASSOPTIONcaptionsoff
  \newpage
\fi

\bibliographystyle{IEEEtran}
\bibliography{references}

\begin{thebibliography}{10}
\providecommand{\url}[1]{#1}
\csname url@samestyle\endcsname
\providecommand{\newblock}{\relax}
\providecommand{\bibinfo}[2]{#2}
\providecommand{\BIBentrySTDinterwordspacing}{\spaceskip=0pt\relax}
\providecommand{\BIBentryALTinterwordstretchfactor}{4}
\providecommand{\BIBentryALTinterwordspacing}{\spaceskip=\fontdimen2\font plus
\BIBentryALTinterwordstretchfactor\fontdimen3\font minus
  \fontdimen4\font\relax}
\providecommand{\BIBforeignlanguage}[2]{{%
\expandafter\ifx\csname l@#1\endcsname\relax
\typeout{** WARNING: IEEEtran.bst: No hyphenation pattern has been}%
\typeout{** loaded for the language `#1'. Using the pattern for}%
\typeout{** the default language instead.}%
\else
\language=\csname l@#1\endcsname
\fi
#2}}
\providecommand{\BIBdecl}{\relax}
\BIBdecl

\bibitem{chodorow2021stock}
G.~Chodorow-Reich, P.~T. Nenov, and A.~Simsek, ``Stock market wealth and the
  real economy: A local labor market approach,'' \emph{American Economic
  Review}, vol. 111, no.~5, pp. 1613--1657, 2021.

\bibitem{chovancova2018changes}
B.~Chovancova, M.~Dorocakova, and V.~Malacka, ``Changes in industrial structure
  of gdp and stock indices also with regard to the industry 4.0,''
  \emph{Business and Economic Horizons (BEH)}, vol.~14, no. 1232-2019-761, pp.
  402--414, 2018.

\bibitem{li2019individualized}
Z.~Li, D.~Yang, L.~Zhao, J.~Bian, T.~Qin, and T.-Y. Liu, ``Individualized
  indicator for all: Stock-wise technical indicator optimization with stock
  embedding,'' in \emph{Proceedings of the 25th ACM SIGKDD}, 2019, pp.
  894--902.

\bibitem{chen2011stock}
C.~R. Chen, J.~D. Diltz, Y.~Huang, and P.~P. Lung, ``Stock and option market
  divergence in the presence of noisy information,'' \emph{Journal of Banking
  \& Finance}, vol.~35, no.~8, pp. 2001--2020, 2011.

\bibitem{jiang2021applications}
W.~Jiang, ``Applications of deep learning in stock market prediction: recent
  progress,'' \emph{Expert Systems with Applications}, vol. 184, p. 115537,
  2021.

\bibitem{CHEN2021853}
F.~Chen, F.~Wu, J.~Xu, G.~Gao, Q.~Ge, and X.-Y. Jing, ``Adaptive deformable
  convolutional network,'' \emph{Neurocomputing}, vol. 453, pp. 853--864, 2021.

\bibitem{ding2020hierarchical}
Q.~Ding, S.~Wu, H.~Sun, J.~Guo, and J.~Guo, ``Hierarchical multi-scale gaussian
  transformer for stock movement prediction,'' in \emph{International Joint
  Conference on Artificial Intelligence}, 2020, pp. 4640--4646.

\bibitem{hochreiter1997long}
S.~Hochreiter and J.~Schmidhuber, ``Long short-term memory,'' \emph{Neural
  Computation}, vol.~9, no.~8, pp. 1735--1780, 1997.

\bibitem{chung2014empirical}
J.~Chung, C.~Gulcehre, K.~Cho, and Y.~Bengio, ``Empirical evaluation of gated
  recurrent neural networks on sequence modeling,'' \emph{arXiv preprint
  arXiv:1412.3555}, 2014.

\bibitem{gupta2022stocknet}
U.~Gupta, V.~Bhattacharjee, and P.~S. Bishnu, ``Stocknet—gru based stock
  index prediction,'' \emph{Expert Systems with Applications}, vol. 207, p.
  117986, 2022.

\bibitem{zhang2017stock}
L.~Zhang, C.~Aggarwal, and G.-J. Qi, ``Stock price prediction via discovering
  multi-frequency trading patterns,'' in \emph{Proceedings of the 23rd ACM
  SIGKDD}, 2017, pp. 2141--2149.

\bibitem{xu2021hist}
W.~Xu, W.~Liu, L.~Wang, Y.~Xia, J.~Bian, J.~Yin, and T.-Y. Liu, ``Hist: A
  graph-based framework for stock trend forecasting via mining concept-oriented
  shared information,'' \emph{arXiv preprint arXiv:2110.13716}, 2021.

\bibitem{casanova2018graph}
P.~V. G. C.~A. Casanova, A.~R.~P. Lio, and Y.~Bengio, ``Graph attention
  networks,'' \emph{ICLR. Petar Velickovic Guillem Cucurull Arantxa Casanova
  Adriana Romero Pietro Li{\`o} and Yoshua Bengio}, 2018.

\bibitem{lin2021learning}
H.~Lin, D.~Zhou, W.~Liu, and J.~Bian, ``Learning multiple stock trading
  patterns with temporal routing adaptor and optimal transport,'' in
  \emph{Proceedings of the 27th ACM SIGKDD}, 2021, pp. 1017--1026.

\bibitem{gavrishchaka2003volatility}
V.~V. Gavrishchaka and S.~B. Ganguli, ``Volatility forecasting from multiscale
  and high-dimensional market data,'' \emph{Neurocomputing}, vol.~55, no. 1-2,
  pp. 285--305, 2003.

\bibitem{kumbure2022machine}
M.~M. Kumbure, C.~Lohrmann, P.~Luukka, and J.~Porras, ``Machine learning
  techniques and data for stock market forecasting: a literature review,''
  \emph{Expert Systems with Applications}, p. 116659, 2022.

\bibitem{li2016stock}
L.~Li, S.~Leng, J.~Yang, and M.~Yu, ``Stock market autoregressive dynamics: a
  multinational comparative study with quantile regression,''
  \emph{Mathematical Problems in Engineering}, vol. 2016, 2016.

\bibitem{ariyo2014stock}
A.~A. Ariyo, A.~O. Adewumi, and C.~K. Ayo, ``Stock price prediction using the
  arima model,'' in \emph{2014 UKSim-AMSS 16th International Conference on
  Computer Modelling and Simulation}.\hskip 1em plus 0.5em minus 0.4em\relax
  IEEE, 2014, pp. 106--112.

\bibitem{wu2006effective}
M.-C. Wu, S.-Y. Lin, and C.-H. Lin, ``An effective application of decision tree
  to stock trading,'' \emph{Expert Systems with Applications}, vol.~31, no.~2,
  pp. 270--274, 2006.

\bibitem{kim2018forecasting}
H.~Y. Kim and C.~H. Won, ``Forecasting the volatility of stock price index: A
  hybrid model integrating lstm with multiple garch-type models,'' \emph{Expert
  Systems with Applications}, vol. 103, pp. 25--37, 2018.

\bibitem{jin2020stock}
Z.~Jin, Y.~Yang, and Y.~Liu, ``Stock closing price prediction based on
  sentiment analysis and lstm,'' \emph{Neural Computing and Applications},
  vol.~32, no.~13, pp. 9713--9729, 2020.

\bibitem{ballings2015evaluating}
M.~Ballings, D.~Van~den Poel, N.~Hespeels, and R.~Gryp, ``Evaluating multiple
  classifiers for stock price direction prediction,'' \emph{Expert Systems with
  Applications}, vol.~42, no.~20, pp. 7046--7056, 2015.

\bibitem{le2019fast}
T.~Le, B.~Vo, H.~Fujita, N.-T. Nguyen, and S.~W. Baik, ``A fast and accurate
  approach for bankruptcy forecasting using squared logistics loss with
  gpu-based extreme gradient boosting,'' \emph{Information Sciences}, vol. 494,
  pp. 294--310, 2019.

\bibitem{kim2000genetic}
K.-j. Kim and I.~Han, ``Genetic algorithms approach to feature discretization
  in artificial neural networks for the prediction of stock price index,''
  \emph{Expert Systems with Applications}, vol.~19, no.~2, pp. 125--132, 2000.

\bibitem{khashei2010artificial}
M.~Khashei and M.~Bijari, ``An artificial neural network (p, d, q) model for
  timeseries forecasting,'' \emph{Expert Systems with Applications}, vol.~37,
  no.~1, pp. 479--489, 2010.

\bibitem{kara2011predicting}
Y.~Kara, M.~A. Boyacioglu, and {\"O}.~K. Baykan, ``Predicting direction of
  stock price index movement using artificial neural networks and support
  vector machines: The sample of the istanbul stock exchange,'' \emph{Expert
  Systems with Applications}, vol.~38, no.~5, pp. 5311--5319, 2011.

\bibitem{cheng2010hybrid}
C.-H. Cheng, T.-L. Chen, and L.-Y. Wei, ``A hybrid model based on rough sets
  theory and genetic algorithms for stock price forecasting,''
  \emph{Information Sciences}, vol. 180, no.~9, pp. 1610--1629, 2010.

\bibitem{aguilar2015genetic}
R.~Aguilar-Rivera, M.~Valenzuela-Rend{\'o}n, and J.~Rodr{\'\i}guez-Ortiz,
  ``Genetic algorithms and darwinian approaches in financial applications: A
  survey,'' \emph{Expert Systems with Applications}, vol.~42, no.~21, pp.
  7684--7697, 2015.

\bibitem{xiaoning2019stock}
C.~Xiaoning, W.~Shang, F.~Jiang, and W.~Shouyang, ``Stock index forecasting by
  hidden markov models with trends recognition,'' in \emph{2019 IEEE
  International Conference on Big Data (BigData)}.\hskip 1em plus 0.5em minus
  0.4em\relax IEEE, 2019, pp. 5292--5297.

\bibitem{lin2013svm}
Y.~Lin, H.~Guo, and J.~Hu, ``An svm-based approach for stock market trend
  prediction,'' in \emph{The 2013 International Joint Conference on Neural
  Networks (IJCNN)}.\hskip 1em plus 0.5em minus 0.4em\relax IEEE, 2013, pp.
  1--7.

\bibitem{li2019multi}
C.~Li, D.~Song, and D.~Tao, ``Multi-task recurrent neural networks and
  higher-order markov random fields for stock price movement prediction:
  Multi-task rnn and higer-order mrfs for stock price classification,'' in
  \emph{Proceedings of the 25th ACM SIGKDD}.\hskip 1em plus 0.5em minus
  0.4em\relax ACM.

\bibitem{zhang2022intraday}
Q.~Zhang, P.~Zhang, and F.~Zhou, ``Intraday and interday features in the
  high-frequency data: Pre-and post-crisis evidence in china’s stock
  market,'' \emph{Expert Systems with Applications}, p. 118321, 2022.

\bibitem{picasso2019technical}
A.~Picasso, S.~Merello, Y.~Ma, L.~Oneto, and E.~Cambria, ``Technical analysis
  and sentiment embeddings for market trend prediction,'' \emph{Expert Systems
  with Applications}, vol. 135, pp. 60--70, 2019.

\bibitem{wu2018hybrid}
H.~Wu, W.~Zhang, W.~Shen, and J.~Wang, ``Hybrid deep sequential modeling for
  social text-driven stock prediction,'' in \emph{Proceedings of the 27th ACM
  International Conference on Information and Knowledge Management}, 2018, pp.
  1627--1630.

\bibitem{zhang2018new}
R.~Zhang, Z.~Yuan, and X.~Shao, ``A new combined cnn-rnn model for sector stock
  price analysis,'' in \emph{2018 IEEE 42nd Annual Computer Software and
  Applications Conference (COMPSAC)}, vol.~2.\hskip 1em plus 0.5em minus
  0.4em\relax IEEE, 2018, pp. 546--551.

\bibitem{chen2015lstm}
K.~Chen, Y.~Zhou, and F.~Dai, ``A lstm-based method for stock returns
  prediction: A case study of china stock market,'' in \emph{2015 IEEE
  International Conference on Big Data (BigData)}.\hskip 1em plus 0.5em minus
  0.4em\relax IEEE, 2015, pp. 2823--2824.

\bibitem{nelson2017stock}
D.~M. Nelson, A.~C. Pereira, and R.~A. De~Oliveira, ``Stock market's price
  movement prediction with lstm neural networks,'' in \emph{2017 International
  Joint Conference on Neural Networks (IJCNN)}.\hskip 1em plus 0.5em minus
  0.4em\relax IEEE, 2017, pp. 1419--1426.

\bibitem{xu2022stock}
H.~Xu, L.~Chai, Z.~Luo, and S.~Li, ``Stock movement prediction via gated
  recurrent unit network based on reinforcement learning with incorporated
  attention mechanisms,'' \emph{Neurocomputing}, vol. 467, pp. 214--228, 2022.

\bibitem{feng2018enhancing}
F.~Feng, H.~Chen, X.~He, J.~Ding, M.~Sun, and T.-S. Chua, ``Enhancing stock
  movement prediction with adversarial training,'' \emph{International Joint
  Conference on Artificial Intelligence}, 2019.

\bibitem{dacorogna1996changing}
M.~M. Dacorogna, C.~L. Gauvreau, U.~A. M{\"u}ller, R.~B. Olsen, and O.~V.
  Pictet, ``Changing time scale for short-term forecasting in financial
  markets,'' \emph{Journal of forecasting}, vol.~15, no.~3, pp. 203--227, 1996.

\bibitem{geva1998scalenet}
A.~B. Geva, ``Scalenet-multiscale neural-network architecture for time series
  prediction,'' \emph{IEEE Transactions on neural networks}, vol.~9, no.~6, pp.
  1471--1482, 1998.

\bibitem{fernandez2019meta}
C.~Fern{\'a}ndez, L.~Salinas, and C.~E. Torres, ``A meta extreme learning
  machine method for forecasting financial time series,'' \emph{Applied
  Intelligence}, vol.~49, no.~2, pp. 532--554, 2019.

\bibitem{liu2020multi}
G.~Liu, Y.~Mao, Q.~Sun, H.~Huang, W.~Gao, X.~Li, J.~Shen, R.~Li, and X.~Wang,
  ``Multi-scale two-way deep neural network for stock trend prediction.'' in
  \emph{International Joint Conference on Artificial Intelligence}, 2020, pp.
  4555--4561.

\bibitem{chen2022multi}
Y.~Chen, F.~Ding, and L.~Zhai, ``Multi-scale temporal features extraction based
  graph convolutional network with attention for multivariate time series
  prediction,'' \emph{Expert Systems with Applications}, vol. 200, p. 117011,
  2022.

\bibitem{xu2022hgnn}
C.~Xu, H.~Huang, X.~Ying, J.~Gao, Z.~Li, P.~Zhang, J.~Xiao, J.~Zhang, and
  J.~Luo, ``Hgnn: Hierarchical graph neural network for predicting the
  classification of price-limit-hitting stocks,'' \emph{Information Sciences},
  vol. 607, pp. 783--798, 2022.

\bibitem{chen2018incorporating}
Y.~Chen, Z.~Wei, and X.~Huang, ``Incorporating corporation relationship via
  graph convolutional neural networks for stock price prediction,'' in
  \emph{Proceedings of the 27th ACM International Conference on Information and
  Knowledge Management}, 2018, pp. 1655--1658.

\bibitem{feng2019temporal}
F.~Feng, X.~He, X.~Wang, C.~Luo, Y.~Liu, and T.-S. Chua, ``Temporal relational
  ranking for stock prediction,'' \emph{ACM Transactions on Information Systems
  (TOIS)}, vol.~37, no.~2, pp. 1--30, 2019.

\bibitem{kim2019hats}
R.~Kim, C.~H. So, M.~Jeong, S.~Lee, J.~Kim, and J.~Kang, ``Hats: A hierarchical
  graph attention network for stock movement prediction,'' \emph{arXiv preprint
  arXiv:1908.07999}, 2019.

\bibitem{yang2020qlib}
X.~Yang, W.~Liu, D.~Zhou, J.~Bian, and T.-Y. Liu, ``Qlib: An ai-oriented
  quantitative investment platform,'' \emph{arXiv preprint arXiv:2009.11189},
  2020.

\bibitem{PatchTST}
Y.~Nie, N.~H.~Nguyen, P.~Sinthong, and J.~Kalagnanam, ``A time series is worth
  64 words: Long-term forecasting with transformers,'' in \emph{International
  Conference on Learning Representations}, 2023.

\bibitem{liu2023itransformer}
Y.~Liu, T.~Hu, H.~Zhang, H.~Wu, S.~Wang, L.~Ma, and M.~Long, ``itransformer:
  Inverted transformers are effective for time series forecasting,''
  \emph{arXiv preprint arXiv:2310.06625}, 2023.

\bibitem{li2024master}
T.~Li, Z.~Liu, Y.~Shen, X.~Wang, H.~Chen, and S.~Huang, ``Master: Market-guided
  stock transformer for stock price forecasting,'' in \emph{Proceedings of the
  AAAI Conference on Artificial Intelligence}, vol.~38, no.~1, 2024, pp.
  162--170.

\bibitem{paszke2019pytorch}
A.~Paszke, S.~Gross, F.~Massa, A.~Lerer, J.~Bradbury, G.~Chanan, T.~Killeen,
  Z.~Lin, N.~Gimelshein, L.~Antiga \emph{et~al.}, ``Pytorch: An imperative
  style, high-performance deep learning library,'' \emph{Advances in neural
  information processing systems}, vol.~32, 2019.

\bibitem{van2008visualizing}
L.~van~der Maaten and G.~Hinton, ``Visualizing data using t-sne,'' \emph{J Mach
  Learn Res}, vol.~9, no.~86, pp. 2579--2605, 2008.

\end{thebibliography}

\begin{IEEEbiography}[{\includegraphics[width=1in,height=1.25in,clip,keepaspectratio]{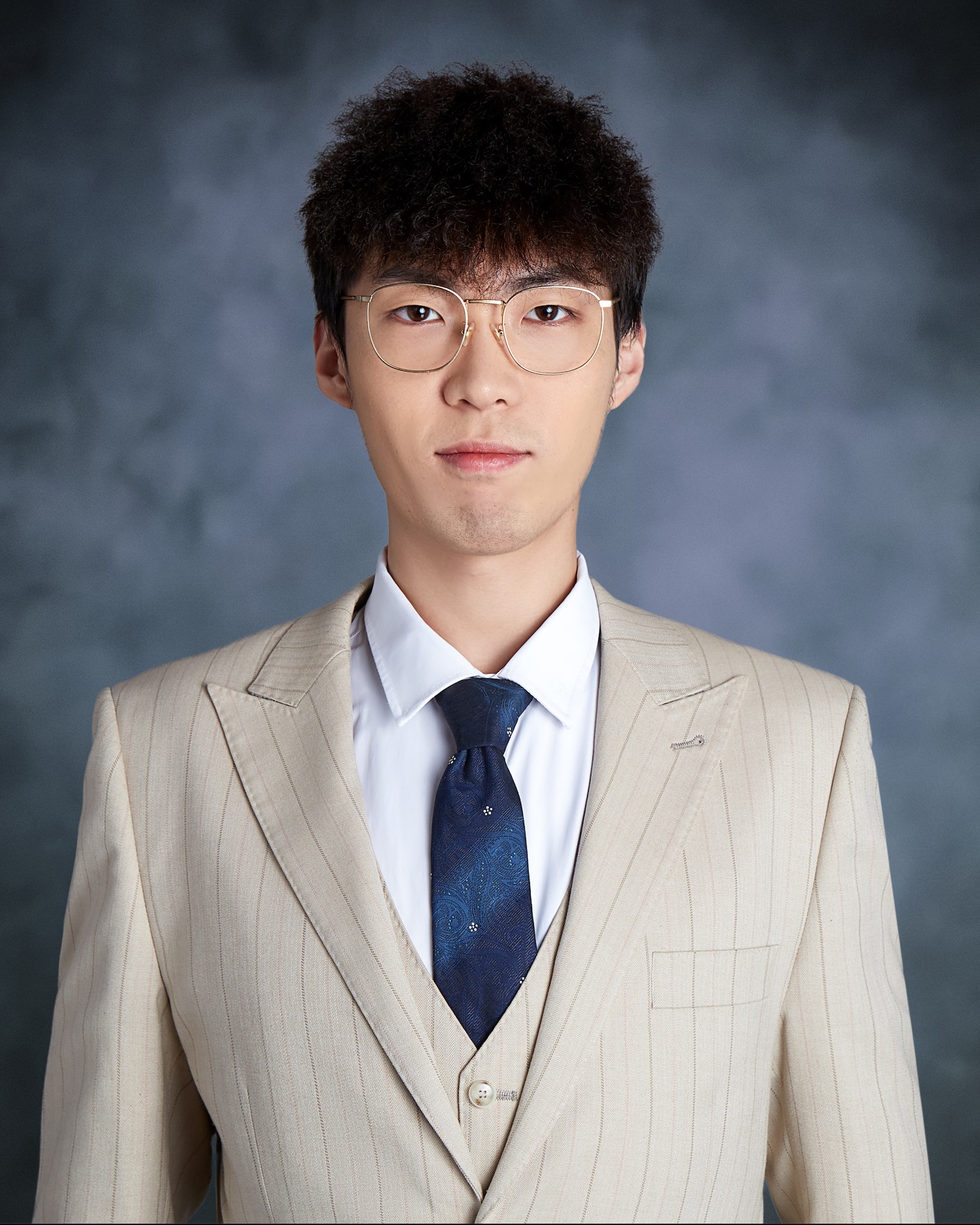}}]{Mingjie Wang}
    received his M.Phil. degree from the Department of Computer Science, BNU-HKBU United International College (UIC), in 2023, and his B.E. degree from the Department of Computer Science and Technology, Longdong University, China, in 2021. He is currently a research assistant at UIC, where his research interests focus on financial time series analysis and federated learning.
\end{IEEEbiography}

\begin{IEEEbiography}[{\includegraphics[width=1in,height=1.25in,clip,keepaspectratio]{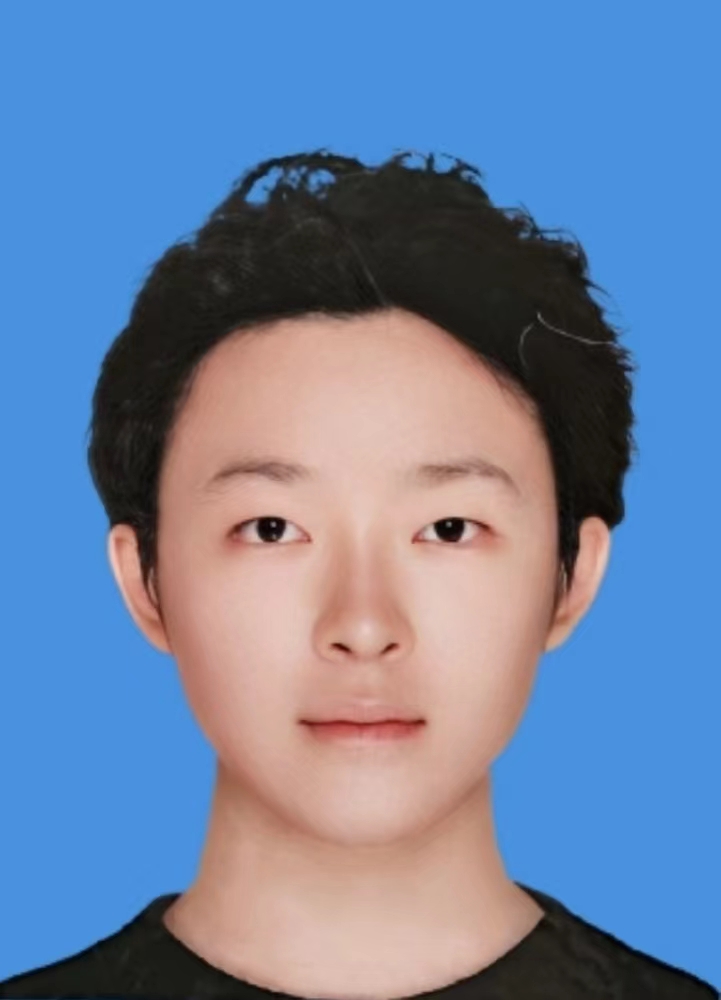}}]{Juanxi Tian}
    is an undergraduate student majoring in Artificial Intelligence in the Department of Computer Science at BNU-HKBU United International College (UIC) since 2021. He has previously served as a guest student at the Shenzhen Institute of Advanced Technology, Chinese Academy of Sciences, and currently works as an intern student at the Artificial Intelligence Research and Innovation Lab, School of Engineering, Westlake University. His current research interests include Neural Network Optimization, CV, Machine Learning and Deep Learning.
\end{IEEEbiography}

\begin{IEEEbiography}[{\includegraphics[width=1in,height=1.25in,clip,keepaspectratio]{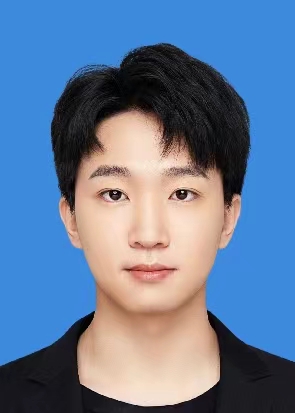}}]{Mingze Zhang}
    is currently an undergraduate student majoring in Computer Science at the School of Engineering and Applied Sciences, Gonzaga University.
\end{IEEEbiography}

\begin{IEEEbiography}[{\includegraphics[width=1in,height=1.25in,clip,keepaspectratio]{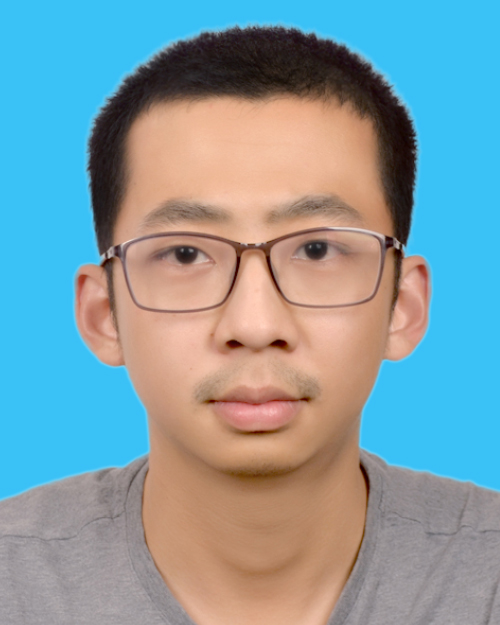}}]{Jianxiong Guo}
	(Member, IEEE) received his Ph.D. degree from the Department of Computer Science, University of Texas at Dallas, USA, in 2021, and his B.E. degree from the School of Chemistry and Chemical Engineering, South China University of Technology, China, in 2015. He is currently an Associate Professor with the Advaned Institute of Natural Sciences, Beijing Normal University, and also with the Guangdong Key Lab of AI and Multi-Modal Data Processing, BNU-HKBU United International College, Zhuhai, China. He is a member of IEEE/ACM/CCF. He has published more than 80 peer-reviewed papers and has been the reviewer for many famous international journals/conferences. His research interests include social networks, wireless sensor networks, combinatorial optimization, and machine learning.
\end{IEEEbiography}

\begin{IEEEbiography}[{\includegraphics[width=1in,height=1.25in,clip,keepaspectratio]{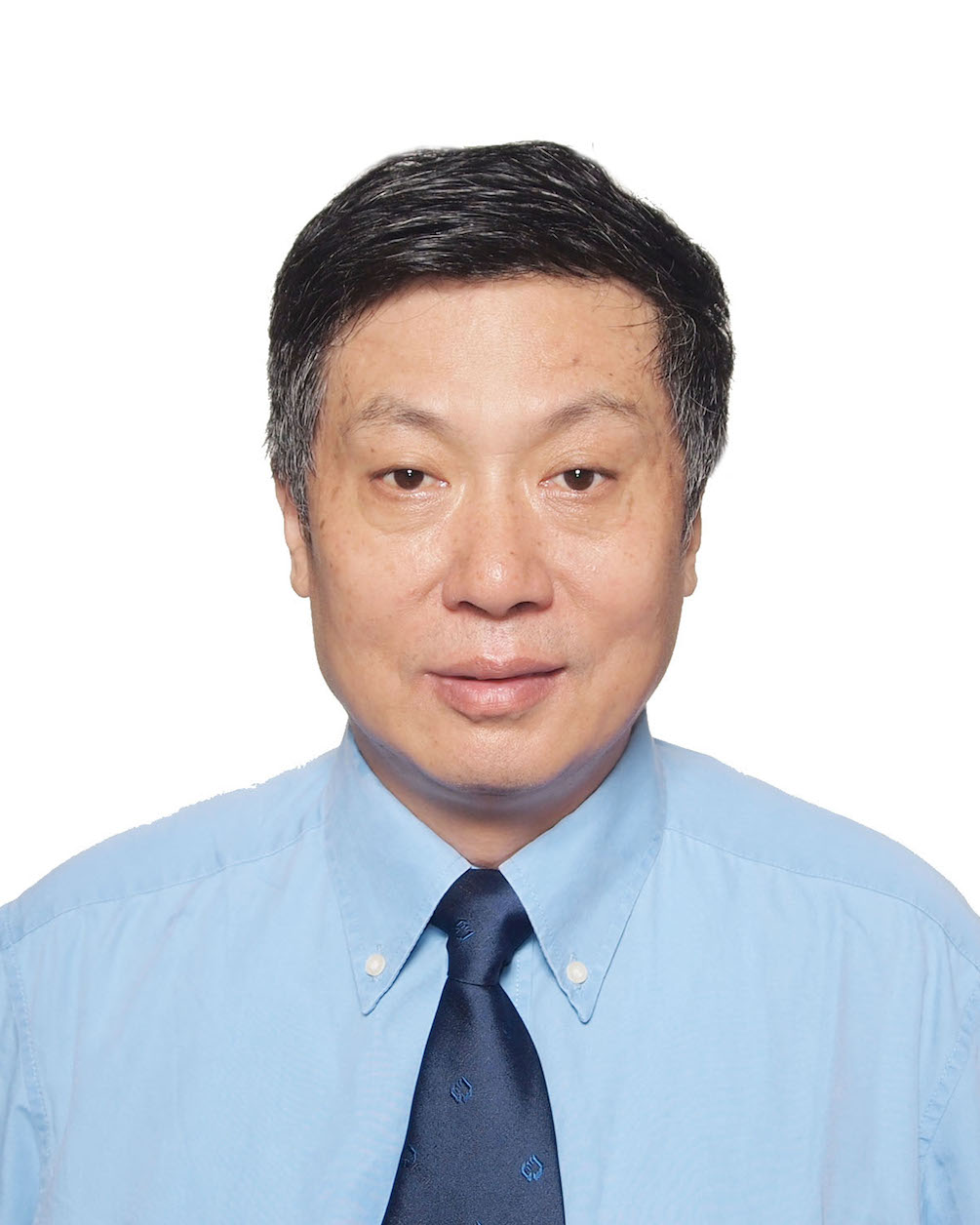}}]{Weijia Jia}
	(Fellow, IEEE) is currently a Chair Professor, Director of the Institute of Artificial Intelligence and Future Networks, Beijing Normal University, and VP for Research at BNU-HKBU United International College (UIC) and has been the Zhiyuan Chair Professor of Shanghai Jiao Tong University, China. He was the Chair Professor and the Deputy Director of State Kay Laboratory of Internet of Things for Smart City at the University of Macau. He received BSc/MSc from Center South University, China in 82/84 and Master of Applied Sci./PhD from Polytechnic Faculty of Mons, Belgium in 92/93, respectively, all in computer science. From 93-95, he joined German National Research Center for Information Science (GMD) in Bonn (St. Augustine) as a research fellow. From 95-13, he worked in City University of Hong Kong as a professor. His contributions have been recognized as optimal network routing and deployment; anycast and QoS routing, sensors networking, AI (knowledge relation extractions; NLP, etc.), and edge computing. He has over 600 publications in the prestige international journals/conferences and research books and book chapters. He has received the best product awards from International Science \& Tech. Expo (Shenzhen) in 20112012 and the 1st Prize of Scientific Research Awards from the Ministry of Education of China in 2017 (list 2). He has served as area editor for various prestige international journals, chair, and PC member/keynote speaker for many top international conferences. He is the Fellow of IEEE and the Distinguished Member of CCF.
\end{IEEEbiography}

\end{document}